\documentclass{aa}  

\usepackage{tabularx}
\usepackage{colortbl}
\usepackage{graphicx}
\usepackage{hyperref}
\usepackage{xcolor}
\usepackage{lscape} 
\usepackage{arydshln} 
\usepackage{multirow}
\usepackage{booktabs} 
\usepackage{float}  
\newcommand{\±}{$\pm$}
\newcommand{\Msun}{ M$_\odot$ }
\newcommand{\rom}[1]{\uppercase\expandafter{\romannumeral #1\relax}}

\usepackage{txfonts}
\begin{document} 


\title{Molecular analysis of a high-mass prestellar core candidate in W43-MM1}

\author{J. Molet \inst{1} \and
          N. Brouillet \inst{1} \and
          T. Nony \inst{2} \and
          A. Gusdorf \inst{3,4} \and
          F. Motte \inst{2,5} \and
          D. Despois \inst{1} \and
          F. Louvet \inst{6} \and
          S. Bontemps \inst{1} \and 
          F. Herpin \inst{1}
          }

\institute{Laboratoire d'astrophysique de Bordeaux, Univ. Bordeaux, CNRS, B18N, all\'ee Geoffroy Saint-Hilaire, 33615 Pessac, France
             \and
             Universit\'e Grenoble Alpes, CNRS, Institut de Plan\'etologie et d'Astrophysique de Grenoble, F-38000 Grenoble, France
             \and
             Laboratoire de Physique de l'\'Ecole normale sup\'erieure, ENS, Universit\'e PSL, CNRS, Sorbonne Universit\'e, Universit\'e Paris-Diderot, Sorbonne Paris Cit\'e, Paris, France
             \and
             LERMA, Observatoire de Paris, PSL Research University, CNRS, Sorbonne Universit\'es, UPMC Univ. Paris 06, 75231 Paris, France
             \and
             AIM Paris-Saclay D\'epartement d'Astrophysique, CEA, CNRS, Univ. Paris Diderot, CEA-Saclay, F-91191 Gif-sur-Yvette Cedex, France
             \and Departamento de Astronomia de Chile, Universidad de Chile, Santiago, Chile \\   
             }

\date{Received March 20, 2019; accepted May 13, 2019}
 
 
\abstract
{High-mass analogues of low-mass prestellar cores are searched for to constrain the models of high-mass star formation. Several high-mass cores, at various evolutionary stages, have been recently identified towards the massive star-forming region W43-MM1 and amongst them a high-mass prestellar core candidate.}
{We aim to characterise the chemistry in this high-mass prestellar core candidate, referred to as W43-MM1 core \#6, and its environment.}
{Using ALMA high-spatial resolution data of W43-MM1, we have studied the molecular content of core \#6 and a neighbouring high-mass protostellar core, referred to as \#3, which is similar in size and mass to core \#6. We first subtracted the continuum emission using a method based on the density distribution of the intensities on each pixel. Then, from the distribution of detected molecules, we identified the molecules centred on the prestellar core candidate (core \#6) and those associated to shocks related to outflows and filament formation. Then we constrained the column densities and temperatures of the molecules detected towards the two cores.}
{While core \#3 appears to contain a hot core with a temperature of about 190\,K, core \#6 seems to have a lower temperature in the range from 20\,K to 90\,K from a rotational diagram analysis. We have considered different source sizes for core \#6 and the comparison of the abundances of the detected molecules towards the core with various interstellar sources shows that it is compatible with a core of size 1000 au with $T = 20-90$ K or a core of size 500 au with $T \sim 80$ K.}
{Core \#6 of W43-MM1 remains one of the best high-mass prestellar core candidates even if we cannot exclude that it is at the very beginning of the protostellar phase of high-mass star formation.}


\maketitle

\section{Introduction}

Understanding high-mass stars ($M>8M_\odot$) is crucial in modern astrophysics since their energy budget towards the interstellar medium is the most important contribution coming from stars. They form in massive dense cores of $\sim$100 M$_\odot$ within $\sim$0.1 pc \citep{motte07, bontemps10}, by processes far less understood than for low-mass stars. Whereas low-mass stars have clear distinct phases of evolution from prestellar cores to young stars, we do not know if a prestellar core phase also exists for high-mass stars. While the UV radiation pressure problem \citep{wolfire87} is mostly solved in most recent 3D modelling, two competitive models remain: the 'core-fed' model \citep[e.g.][]{mckee03} and the 'clump-fed' model \citep[e.g.][]{bonnell06, smith09, vazquez17}. The first model supposes the existence of starless massive dense cores, whereas, for the second one, low-mass prestellar cores can become high-mass protostars by collecting surrounding gas. The observation of a high-mass prestellar core is challenging due to the lower number of high-mass stars compared to the number of low-mass stars (only $\sim$1\% of the Galactic stellar population) and the short duration of the hypothetical prestellar phase \cite[$\sim$$10^4-10^5$ yr; ][and references therein] {tige17}. \cite{motte18r} and \cite{louvet18r} list the current high-mass prestellar core candidates: CygX-N53-MM2 \citep{duartecabral14}, G11.92-0.61-MM2 \citep{cyganowski14}, G11.11-P6-SMA1 \citep{wang14}, G028CA9 \citep{kong17}, and W43-MM1\#6 \citep{nony18}. No detailed study about their molecular content has been done so far. We investigate here the latter one, W43-MM1\#6.

The W43-MM1 ridge is a massive molecular cloud of  $2\times10^4$\,M$_\odot$ contained in 6\,pc$^2$. It is the main submillimeter fragment of the W43 complex \citep{motte03, nguyen13}, located at the end of the Galactic bar at a distance of 5.5\,kpc from the Sun \citep{nguyen11, zhang14}. W43-MM1 is associated to OH, CH$_3$OH,  and H$_{2}$O masers \citep[respectively][]{braz83, walsh98, valdettaro01}, which are high-mass stellar activity signposts \citep{motte03}. At 0.05 pc resolution, W43-MM1 appears as a mini-starburst cluster with a high star formation rate of 6000 M$_\odot$.Myr$^{-1}$ and N1a as the most massive dense core \citep{louvet14}. \cite{louvet16} identified a large-scale shock ($\sim$5 pc) along the filament, consistent with numerical models of a collision of two giant molecular clouds  \citep[e.g. ][]{wu17}, and intermediate-scale shocks ($\sim$0.2 pc) tracing protostellar outflows. The presence of small-scale shocks due to the collision of gas inflows on the cores has been suggested but not observed yet probably because of the limited sensitivity and/or angular resolution. More recently, \cite{motte18} identified a large population of cores with $\sim$2000 au typical sizes, and numerous high-mass cores. This suggests that W43-MM1 is one of the youngest and richest clusters of high-mass cores in the Milky Way. This sample is thus excellent to search for a high-mass prestellar core \citep{nony18}.  

Two neighbouring cores with the same size and mass have been studied by \cite{nony18}. The first core, identified as core \#3, has clear outflows traced by SiO(5-4) and CO(2-1) transitions and a line forest seen over the continuum band, which qualifies it as a protostellar object. Core \#6 in contrast does not show any outflow and just a few lines are seen in the continuum band; it is thus a good high-mass prestellar core candidate. During the evolution from the molecular cloud to the formation of the protostar, the composition of the gas evolves \citep[e.g.][]{gerner14}. The formation and destruction of molecules depends on the density of the environment, temperature, time, and dynamics. A rise in temperature will allow molecules to desorb from the grain mantles to the gas phase and thus to be detectable by rotational spectroscopy. Complex organic molecules (COMs) have already been detected in both high-mass and low-mass star-forming regions. The typical radius of hot corinos, where the dust temperature is 100 K, is <100 au, while it is a few thousands au for hot cores \citep[e.g.][]{herbst09}. We propose here to determine the nature of the W43-MM1 high-mass core \#6, constraining the physical and chemical conditions of the core from the observed abundances and excitation temperatures of the detected molecules. 

The article is organised as follows. We present the data and the method used to subtract the continuum in Sect. 2. In Sect. 3, we analyse the molecular content of cores \#3 and \#6. We study the distribution of the molecules to select those that are directly related to core \#6 and we estimate their temperature and column densities. In Sect. 4 we compare our results with other star-forming regions and discuss the nature of cores \#3 and \#6.

\section{Observations and data reduction}

\subsection{Data set}
Observations were carried out in Cycle 2 (project \#2013.1.01365.S) and Cycle 3 (\#2015.1.01273.S), with the ALMA 12 m and ACA 7 m arrays. W43-MM1 was imaged in band 6 (between 216 and 234 GHz) with a 78"$\times$53" (2.1 pc$\times$1.4 pc) mosaic composed of 33 fields with the 12 m array and 11 fields with ACA. The total bandwidth is 4.8 GHz made of two continuum bands and seven narrow bands centred on lines of interest, with a spectral resolution ranging from 122 to 976 kHz (0.17 to 1.26 km.s$^{-1}$) and a spatial resolution of $\sim$0.44" ($\sim$2400 au) (see Table \ref{table:bands}). The continuum band centred on 232.4 GHz is taken from Cycle 3 data and has a lower spatial resolution ($\sim$0.57" or $\sim$3200 au). Data reduction was made with CASA 4.7.2. \citep{mcmullin07}. The cleaning was performed with a 0.5 Briggs robustness weighting, using the multiscale option and with a threshold of $\sim$1$\sigma$ cropped on a region excluding the borders of the mosaic to avoid problems of divergence. 

\begin{table}[hbt!]
        \setlength{\tabcolsep}{4pt}
      \caption[]{Parameters of the ALMA spectral windows.}
      \label{table:bands}
      \begin{tabular}{lccccc}
            \hline
            \hline
            Spectral    & $\nu_{obs}$   & Bandwidth     & \multicolumn{2}{c}{Resolution}        & rms \\
            windows     & [GHz]         & [MHz]         & [$\arcsec$ $\times$ $\arcsec$]              & [km.s$^{-1}$] &[K]  \\  
            \hline
            OC$^{33}$S & 216.076        & 234           & 0.55 $\times$ 0.40                 & 0.17                  & 0.37 \\
            SiO                 & 217.033       & 234           & 0.54 $\times$ 0.39            & 0.34                  & 0.42   \\
            H$_2$CO     & 218.150       & 234           & 0.54 $\times$ 0.39            & 0.17                    & 0.38 \\
            C$^{18}$O   & 219.488       & 117           & 0.53 $\times$ 0.39            & 0.17                    & 0.49 \\
            SO          & 219.877       & 117           & 0.54 $\times$ 0.39            & 0.33                    & 0.44 \\       
            CO          & 230.462       & 469           & 0.52 $\times$ 0.38            & 1.27                    & 0.25 \\
            $^{13}$CS   & 231.144       & 469           & 0.52 $\times$ 0.37            & 0.32                    & 0.28\\
            Cont. 1 $^ a$& 232.4        & 1875          & 0.66 $\times$ 0.50            & 1.26                    & 0.12\\
            Cont. 2             & 233.4         & 1875          & 0.51 $\times$ 0.36            & 1.26                  & 0.22\\
            \hline
         \end{tabular}
          $^ a$ spectral window observed in Cycle 3. The other windows were observed in Cycle 2.
\end{table}

\begin{table*}[t]
        \small\centering
        \caption[]{Characteristics of the cores located at the tip of the main filament \citep[from][]{motte18}.}
        \label{table:sources}
         \begin{tabular}{cccccccc}
            \hline
            \hline
            Core        & RA                                                    & Dec                                                     & Size                  & FWHM            & M$_{\rm core}$        & n$_{\rm H2}$          & T$_{\rm dust}$\\
            \#          & [J2000]                                                       & [J2000]                                                 & ["$\times$"]          & [au]            & [M$_\odot$]           & [x10$^9$cm$^{-3}$]    &[K] \\
            \hline
            3           & 18$^{\rm h}$47$^{\rm m}$46$\fs$37     & -1$\degr$54$\arcmin$33$\farcs$41      & 0.52$\times$0.47        & 1200          &  59$\pm$2             & 7.6$\pm$0.3                 & 45$\pm$1\\
            6           & 18$^{\rm h}$47$^{\rm m}$46$\fs$16     & -1$\degr$54$\arcmin$33$\farcs$30      & 0.55$\times$0.45        & 1300          &  56$\pm$9             & 6.6$\pm$0.1                 & 23$\pm$2\\
            9           & 18$^{\rm h}$47$^{\rm m}$46$\fs$48     & -1$\degr$54$\arcmin$32$\farcs$54      & 0.63$\times$0.45        & 1600          &  17.8$\pm$0.9 & 1.0$\pm$0.1                   & 50$\pm$1\\
            18          & 18$^{\rm h}$47$^{\rm m}$46$\fs$25     & -1$\degr$54$\arcmin$33$\farcs$41      & 0.93$\times$0.45        & 2600          &  28$\pm$4             & 0.4$\pm$0.1                   & 23$\pm$2\\
            \hline
         \end{tabular}
\end{table*}
 
\subsection{Continuum subtraction}
        Because of the sensitivity of the current instruments, it is now difficult to find channels with no emission line over all the map to subtract the continuum in a data cube. In W43-MM1, \cite{motte18} identify 131 continuum cores with masses ranging from $\sim$1\Msun to $\sim$100 M$_\odot$ and line contamination between 0\% and 74\%. The analysis of the four bright methyl formate (CH$_3$OCHO) lines detected in the 216 GHz band gives a velocity ranging from 94 to 102 km.s$^{-1}$ for the 11 main continuum cores. Due to the numerous lines in the various cores and the different velocities of the cores, line emission is present in almost all channels. It is thus necessary to use a method of continuum subtraction that handles each pixel independently. We developed a method similar to the one presented in \cite{jorgensen16}, using the density distribution of the channel intensities on each pixel. The profile of the distribution is composed of a Gaussian part due to the noise, and a tail towards high intensities associated with the molecular emission. The Gaussian noise distribution peaks on the true continuum value, but the peak of the observed distribution can be displaced due to the contribution of the tail. The best way to obtain a correct continuum value with the density profile method is discussed in \cite{sanchezmonge17}, and the corrected sigma-clipping method (cSCM) seems to be the most effective using their STATCONT algorithm. To properly treat the statistical behaviour of the tail, we use the exponentially modified Gaussian (EMG) function \citep{grushka72}, which is the convolution between a Gaussian and an exponential decay, 
 
        \begin{equation}
                f_{EMG}(x) = A \sigma \lambda \sqrt{\frac{\pi}{2}} {\rm \it exp} [\lambda \frac{2\mu + \lambda \sigma^2 - 2x}{2}] {\rm \it erfc} (\frac{\mu+\lambda\sigma^2-x}{\sqrt{2}\sigma}) ,
        \end{equation}

where A is the peak amplitude, $\sigma$ the variance of the Gaussian, $\lambda$ the relaxation rate of the exponential modifier, $\mu$ the centre of the Gaussian, and ${\rm \it erfc}$ the complementary error function. 

\begin{figure}
                \includegraphics[width=\linewidth]{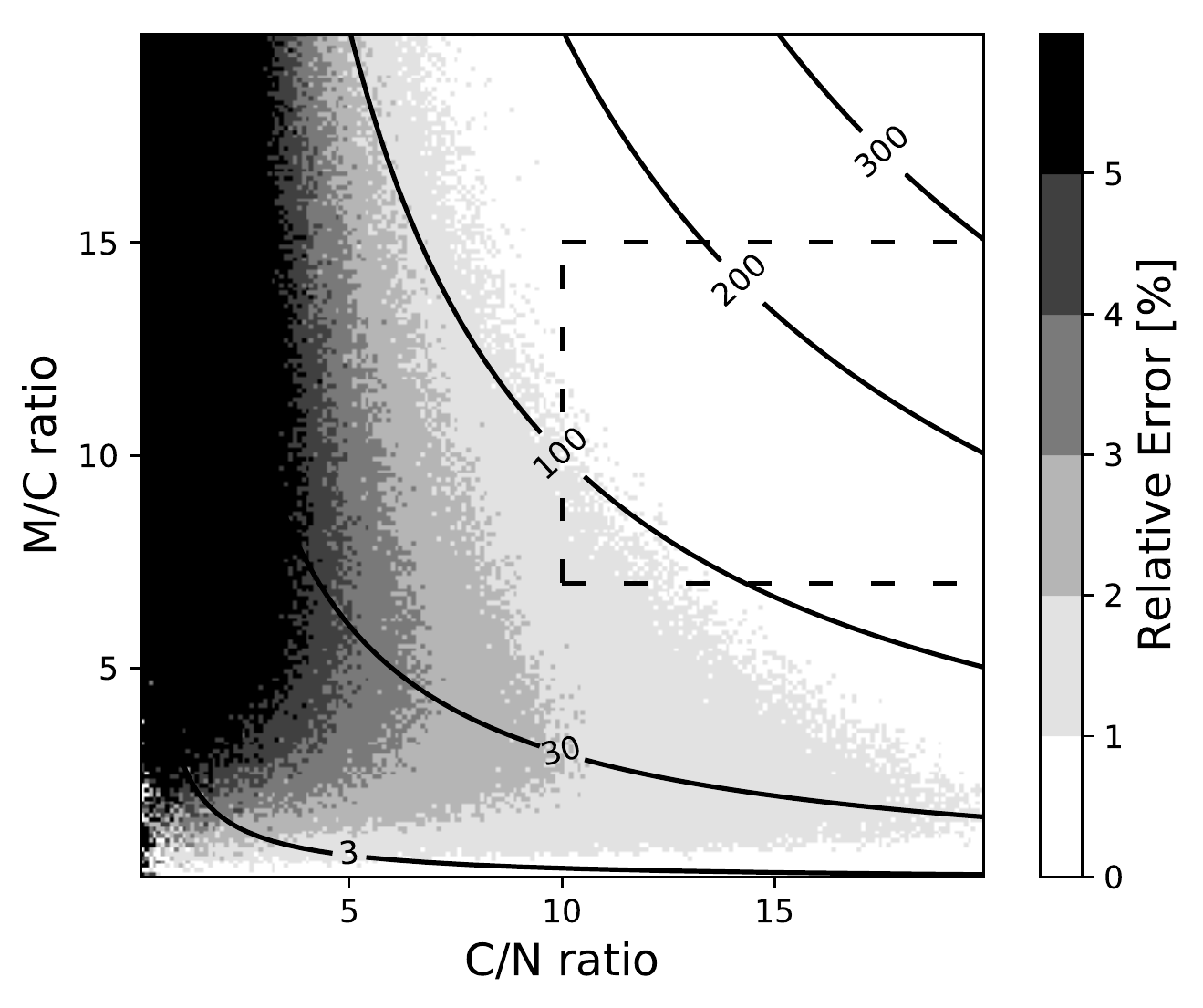}
                \caption{Estimation of the relative error on the continuum level determination, depending on the molecular/continuum (M/C) ratio and continuum/noise (C/N) ratio. The black lines represent the M/N ratio. The typical ratios from the core \#3 pixels are included in the dashed box. Each pixel is the average of the errors obtained from 20 synthetic spectra of 1000 channels with 100 random molecular Gaussian lines that follow a power law comparable to the observed data.}
                \label{fig:error}
        \end{figure}

We compared our EMG method with the cSCM from STATCONT, using synthetic spectra with parameters comparable to our data. Our method is about 20 times faster than STATCONT and we estimated a relative error on the continuum of $\pm$2\% with the EMG method fit versus $\pm$5\% for the cSCM. Figure \ref{fig:error} presents the estimation of the error depending on the following parameters: the continuum level, the maximum of the molecular emission and the noise level. We find that the EMG method is well adapted for this kind of molecular distribution. Part of the derived continuum map is presented in Fig. \ref{fig:Mdecentred}. Comparing with the line-contaminated continuum map of \cite{motte18}, derived from the integration of the full continuum band without getting rid of molecular lines, the positions of the cores are the same but the signal on this map is overestimated on average by $\sim$20\% on continuum cores and up to 60\% for cores \#1 and \#4. As for the line-free continuum map of \cite{motte18}, it is consistent, within 10\%, with the EMG continuum image but it is 1.2 times noisier. 

\section{Results and analysis}

        Out of the 13 high-mass cores with ${\rm M} >16$ M$_\odot$, identified at 1.3mm by \cite{motte18} in the W43-MM1 protocluster, we focus in this paper on the high-mass cores at the south-western tip of the main filament \citep[see Fig. 1 of][]{nony18} and we follow the same nomenclature. Table \ref{table:sources} lists the main characteristics of the four high-mass cores identified in the continuum map (lower right panel in Fig. \ref{fig:Mdecentred}). The sizes and shapes of the cores are the same as in \cite{motte18} and \cite{nony18} and do not change in this frequency range. We are interested in particular in the comparison of cores \#3 and \#6, which are close (0.08 pc, 16000 au), with a similar size of 1200 au and mass of $\sim$60\,M$_\odot$. We study here their spectra, averaged spatially over the size of the continuum cores.

   \subsection{Molecular content of cores \#3 and \#6}
   
        Whereas core \#3 displays a forest of lines, much fewer lines are observed towards core \#6 (see Figs. \ref{fig:Spectra-core3} and \ref{fig:Spectra-core6}, respectively). We first studied the molecular content of core \#3. The 25 molecules that we identified in this core, including isotopologues, are listed in Table \ref{table:MoleculeProp}. Out of these molecules, 17 are identified in core \#6. The line parameters of the 33 detected transitions towards core \#6 are listed in Table \ref{table:LinesC6}. 
        
        Complex molecules are clearly present in core \#6 with the detection of methanol (CH$_3$OH) and its isotopologue ($^{13}$CH$_3$OH), acetaldehyde (CH$_3$CHO), dimethyl ether (CH$_3$OCH$_3$), and methyl formate (CH$_3$OCHO). There is no clear evidence of the presence of other COMs in core \#6, except a tentative detection of the methanol isotopologue CH$_3$$^{18}$OH and of formamide (NH$_2$CHO) that only appear when the corresponding lines are stacked. For CH$_3$$^{18}$OH, eight lines are considered as not contaminated by other species in core \#3 and were selected for core \#6. In the same way, five lines were selected for NH$_2$CHO.
Some of the molecules detected towards core \#6 seem to be directly associated with the continuum core, whereas the others do not peak at the centre of the core and are being interpreted as part of the large-scale features. 

\begin{figure*}[t!]
        \centering
        \includegraphics[width=0.99\linewidth]{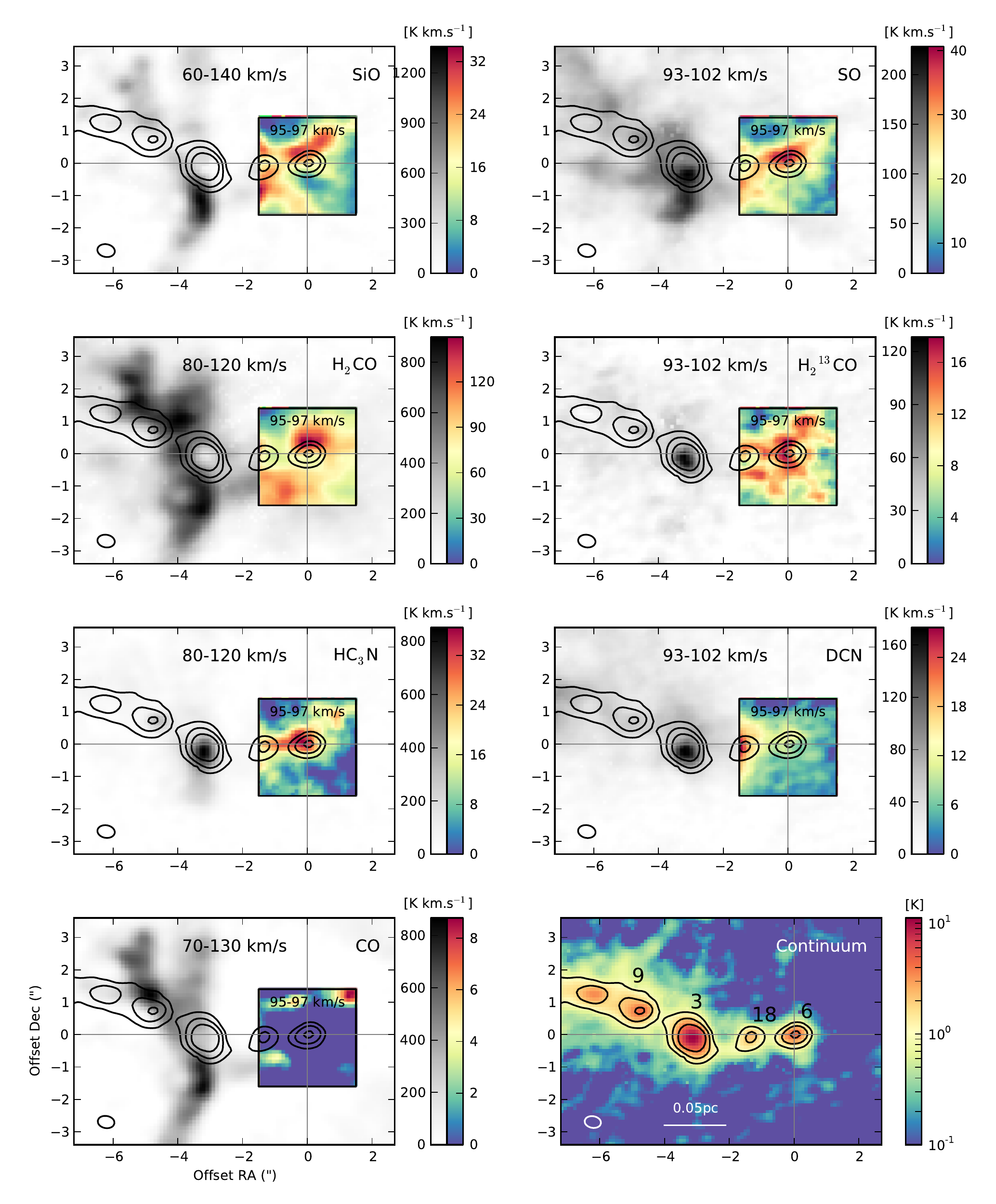}
        \caption{Velocity integrated maps of the molecules that do not peak on core \#6. The velocity range is indicated on the top of each box. For each molecule, grey scale images correspond to the large-scale map with large dynamics in intensity, while the colour maps focus on a narrower region around core \#6 with smaller dynamics in intensity (as indicated by the two wedges on the right side of each panel). Temperature is the main beam brightness temperature $T_{\rm mb}$. The continuum map obtained from the Cont. 2 spectral window data is plotted in the bottom right panel. The contours represent the 5-10-20$\sigma$ continuum emission, with 5$\sigma\approx$ 9mJy/beam. The position is relative to the centre of the continuum core \#6. The beam is represented in the bottom left corner of each map. The CO is completely filtered in the 95-97 km/s velocity range around core \#6 (see Fig. \ref{fig:Spectra-core6}), but its presence is confirmed by the C$^{18}$O emission.}
        \label{fig:Mdecentred}
\end{figure*}

\paragraph{- Molecules not directly associated to core \#6} \hspace{0pt} \\
        For seven of the species detected in core \#6, the large-scale view shows that the distribution of these molecules is not only centred on the continuum cores (see Fig. \ref{fig:Mdecentred}). On small scales, those molecules clearly peak at the north of core \#6 (e.g. SiO) or belong to the environment (see DCN map). On large scales, CO and SiO mainly trace the outflows from core \#3 and \#9, which have been already analysed by \cite{nony18}. SO, H$_2^{13}$CO, and DCN follow the filament observed in continuum emission.

\begin{figure*}[t!]
        \centering
        \includegraphics[width=0.99\linewidth]{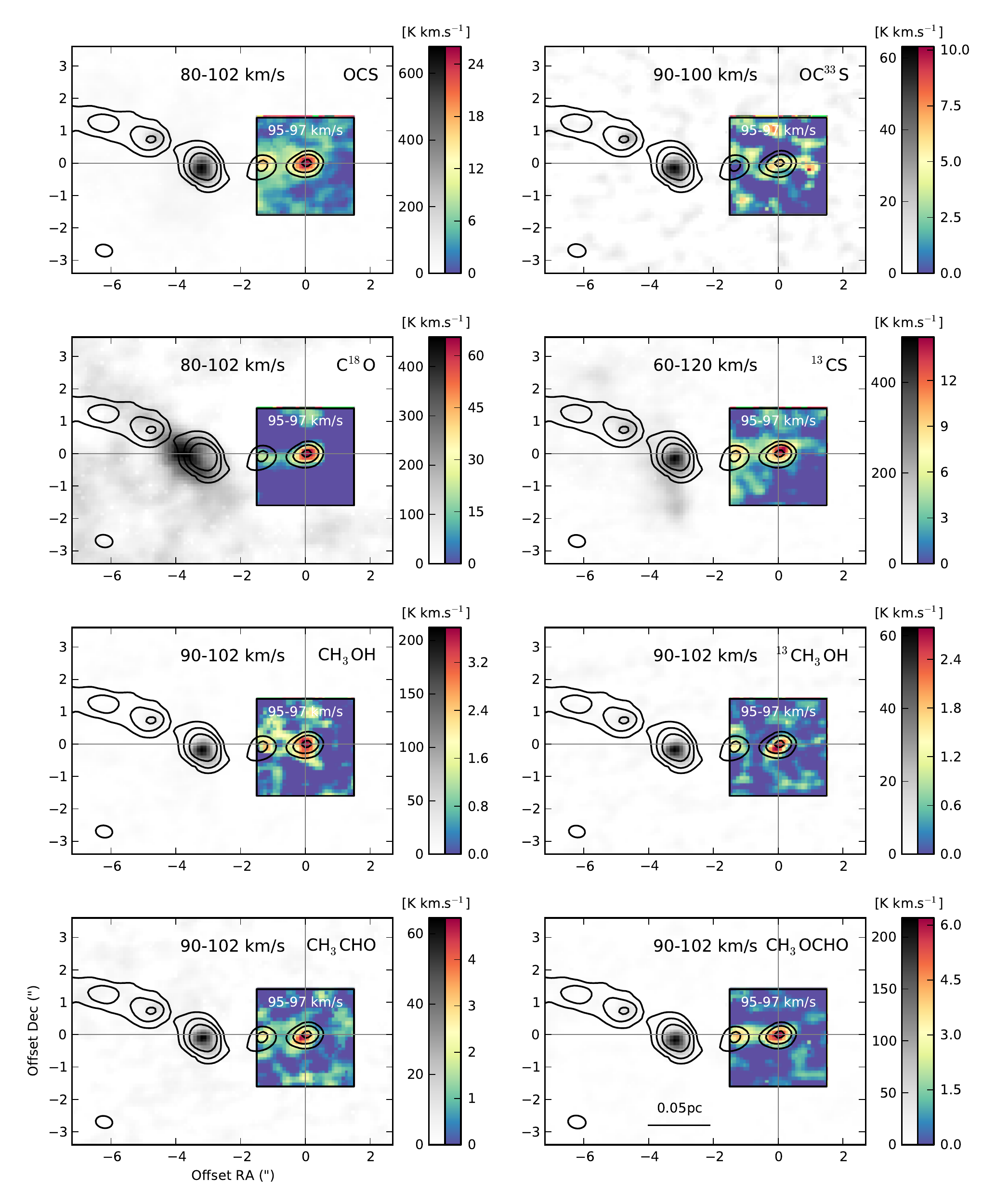}
        \caption{Same as Fig. \ref{fig:Mdecentred} but for molecules centred on core \#6. \newline \newline \newline}
        \label{fig:Mcentred369}
\end{figure*}

\begin{figure*}
        \centering
        \includegraphics[width=\linewidth]{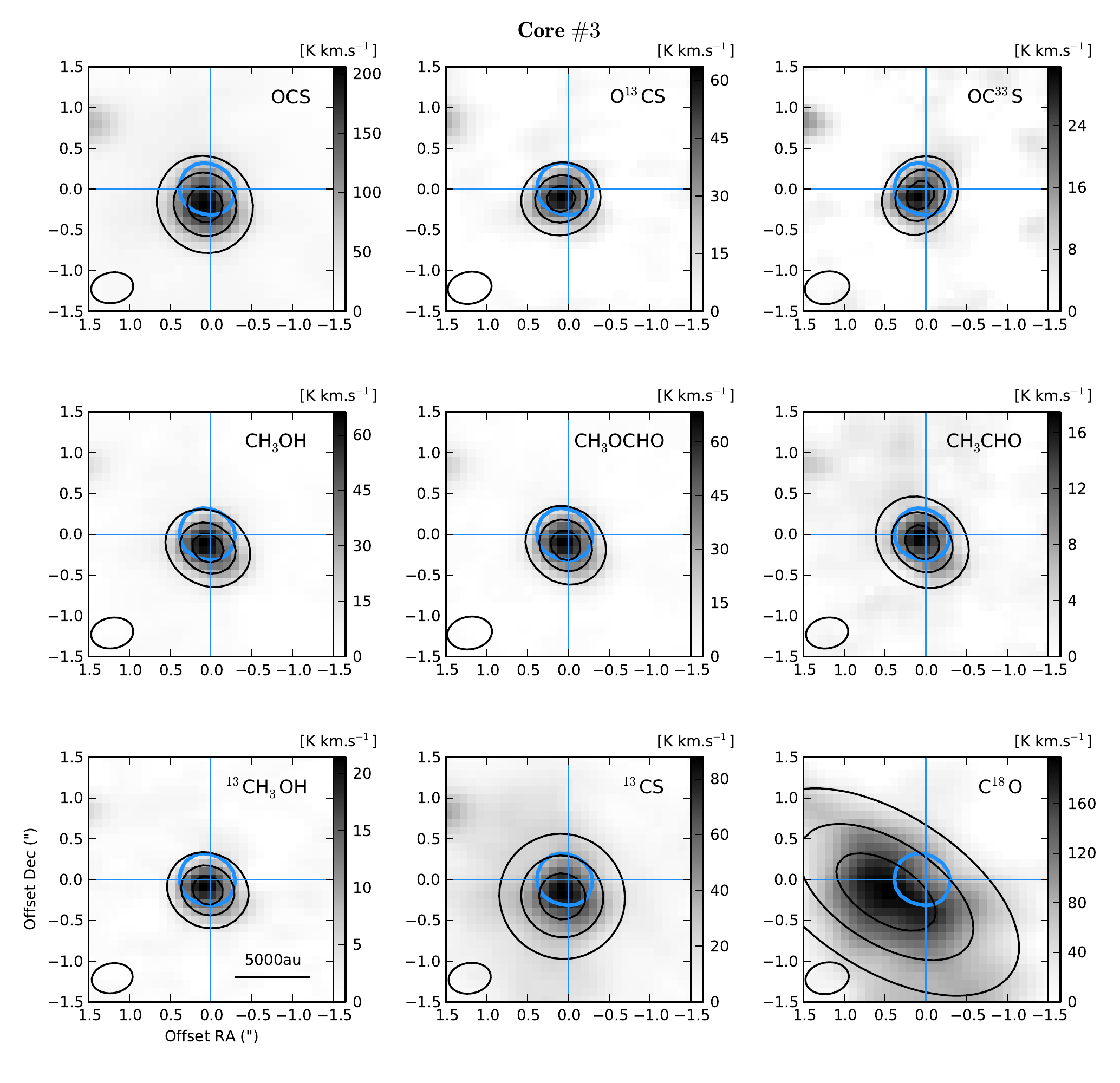}
        \caption{Velocity integrated intensity map (grey scale) around core \#3 for the molecules centred on the continuum core. The velocity range chosen for the integration is 95-97 km/s, which corresponds to the central emission peak. Temperature is the main beam brightness temperature $T_{\rm mb}$. The black contours represent the result of a Gaussian fit to the emission; the levels are 20, 50, and 80\% of the maximum. The maps are centred on the continuum core represented in blue contours (50\% of the maximum). The beam is represented on the bottom left corner of each map.}
        \label{fig:Mcentred3}
\end{figure*}

\begin{figure*}
        \centering
        \includegraphics[width=\linewidth]{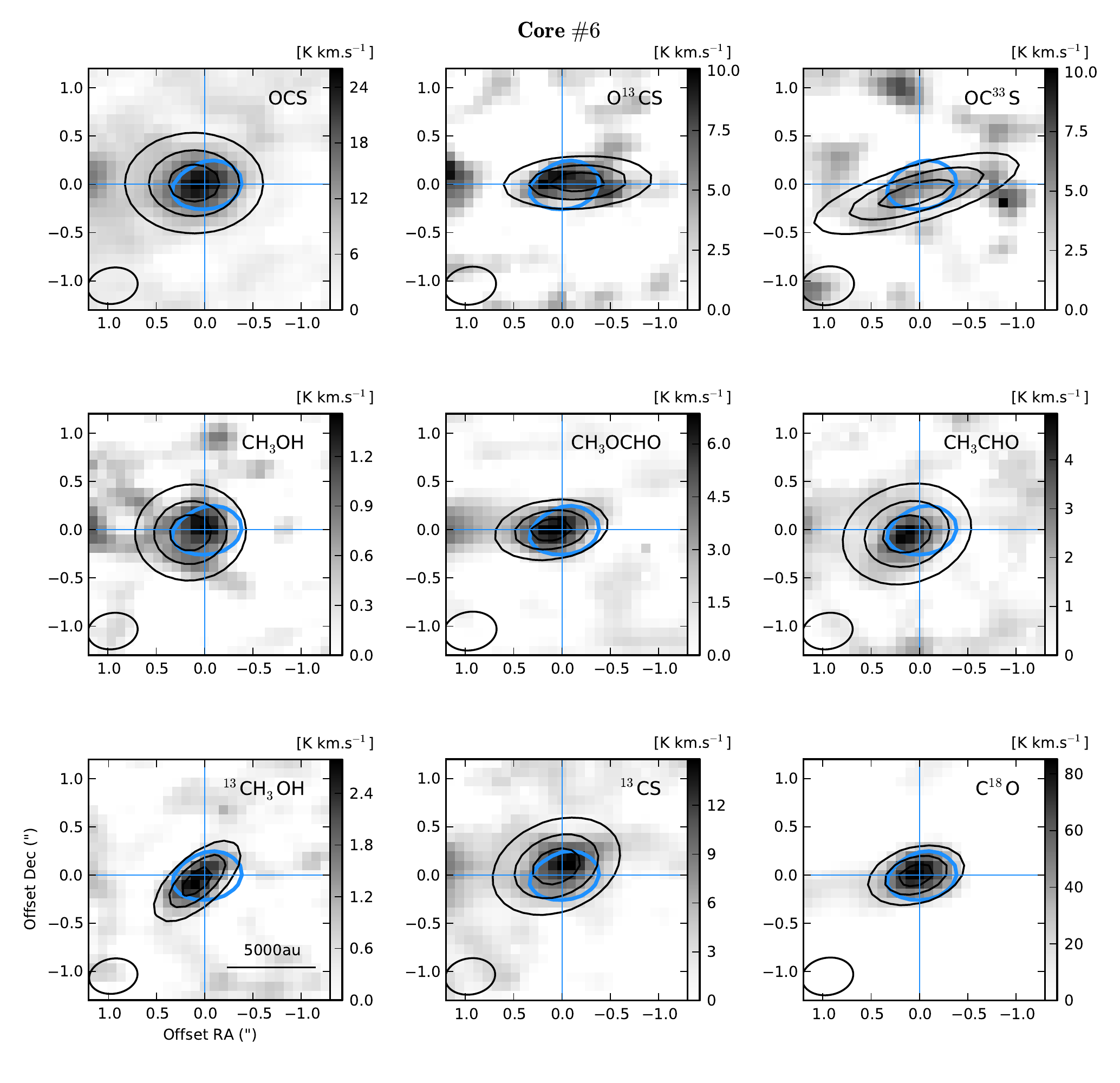}
        \caption{Same as Fig. \ref{fig:Mcentred3} but for source \#6.}
        \label{fig:Mcentred}
\end{figure*}

\paragraph{- Molecules centred on core \#6} \hspace{0pt} \\
        The other ten identified species are clearly centred on the continuum core (see Fig. \ref{fig:Mcentred369}). This suggests that part of the emission of these molecules is not associated to the large-scale shocks expected in a cloud forming though collision or along protostellar outflows, but is directly linked to the presence of the core. The distribution of these molecules in the neighbouring core \#3 is also centred on the continuum peak as shown in Fig. \ref{fig:Mcentred3}. We used a Gaussian fit to derive a deconvolved size of the molecular emission, which appears to be typically close to the size of the beam for all ten molecules observed in core \#6. The results are comparable for the two cores (see Table \ref{table:MoleculeProp}), except for the C$^{18}$O emission, which is three times more extended than the continuum extension of core \#3. We note a recurrent small offset (<0.2", i.e. less than half of the beam) of the molecular emission with respect to the continuum peak, towards the south for core \#3 and towards the east for core \#6 (see Fig. \ref{fig:Mcentred}).  
        
\begin{table*}  
\setlength\tabcolsep{1mm}

\caption{Excitation temperatures and column densities of the detected molecules towards cores \#3 and \#6 derived for a core of the beam size.}

\begin{tabular}{lccccccccccc} 
	\hline\hline
					& \multicolumn{5}{c}{Core \#3}	& 	& \multicolumn{5}{c}{Core \#6}			\\
	\cmidrule(lr){2-6} \cmidrule(lr){8-12} 								
	\multirow{2}{*}{Molecule}	& $\overline{v}$		& $\overline{dv}$			& $N_{\rm tot}$	& $T_{\rm ex}$	& Size		&	& $\overline{v}$		& $\overline{dv}$			& $N_{\rm tot}$	& $T_{\rm ex}$ 		& Size   	\\	
					& [km.s$^{-1}$]		& [km.s$^{-1}$]		& [cm$^{-2}$]			& [K]			& [$\arcsec$ $\times$ $\arcsec$]		&	& [km.s$^{-1}$]		& [km.s$^{-1}$]		& [cm$^{-2}$]		& [K]			& [$\arcsec$ $\times$ $\arcsec$]		\\
	\hline
	C$^{18}$O		& -			& -			& >1.8(17) *		& 190		& 2.36$\times$1.17 	&	& -			& -			&-		& -	& 0.53$\times$0.18	\\
	CH$_3$OH		&97.1$\pm$0.1&5.9$\pm$0.3	& 4.7\±0.4(17)		& 320\±80	& 0.55$\times$0.35	&	&95.8$\pm$0.2&3.3$\pm$0.4	& 4.2\±4.0(17) & 55\±35		& 0.54$\times$0.52 	\\
	$^{13}$CH$_3$OH	&97.0$\pm$0.1&5.3$\pm$0.3	& 9.8\±2.1(16)		& 210\±40	& 0.52$\times$0.36	&	&96.5$\pm$0.2&5.2$\pm$0.5	& -		& -	& <0.55$\times$0.45	\\
	CH$_3$CHO		&96.6$\pm$0.2&5.6$\pm$0.7	& 5.7\±0.1(15)		& 150\±40	& 0.68$\times$0.46 	&	&95.6$\pm$0.1&2.6$\pm$0.4	& 2.6\±1.9(15) & 30\±15		& 0.71$\times$0.54	\\
	$^{13}$CS		&97.0$\pm$0.1&5.0$\pm$0.1	& 8.9(14)			& 190		& 0.94$\times$0.86	&	&95.4$\pm$0.1&3.4$\pm$0.3	&- 		& -	& 0.73$\times$0.48 		\\
	CH$_3$OCH$_3$	&96.3$\pm$0.1&6.7$\pm$0.3	& 9.3\±1.4(16)		& 140\±25	& 0.61$\times$0.53	&	&95.9$\pm$0.2&2.3$\pm$0.7	& - 		& -		& -			\\
	CH$_3$OCHO		&97.3$\pm$0.3&4.7$\pm$0.8	& 1.0\±0.3(17)		& 250\±50	& 0.51$\times$0.28	&	&95.7$\pm$0.2&3.0$\pm$0.3	& 7.0\±4.0(15) & 50\±30	& 0.62$\times$0.22	\\
	OCS				&96.5$\pm$0.1&5.7$\pm$0.1&  2.2(16)			& 190		& 0.70$\times$0.55	&	&95.7$\pm$0.1&3.7$\pm$0.2	& - 		& -		& 0.78$\times$0.56	\\
	O$^{13}$CS		&97.1$\pm$0.1&4.0$\pm$0.1	& 3.3(15)			& 190		& 0.43$\times$0.34	&	&95.1$\pm$0.2&2.1$\pm$0.4	& - 		& -		& <0.55$\times$0.45	\\
	OC$^{33}$S		&96.6$\pm$0.1&3.7$\pm$0.2	& 1.4(15)			& 190		& 0.49$\times$0.24	&	&95.4$\pm$0.3&2.1$\pm$0.6	& - 		& -		& <2.1$\times$0.22	\\
	\hdashline
	SiO $v=0$		& -			& -			& >5.6(14) *		& 190		&	-		&	&95.8$\pm$0.3& 5.1$\pm$0.7& -		& -		& \multirow{7}{*}{\centering Not centred} \\
	SO				&97.0$\pm$0.1&5.6$\pm$0.2& 4.1(15)			& 190		&	-		&	&96.1$\pm$0.1&3.2$\pm$0.3	& - 		& -		&  \\
	H$_2$CO			& -			& -			& >1.0(16) *		& 190		&	-		&	& -			& -			& -	& -		&  \\
	H$_2^{13}$CO	&97.2$\pm$0.1&4.6$\pm$0.2	& 2.1(15)			& 190		&	-		&	&96.0$\pm$0.2&2.4$\pm$0.6	& - 		& -		&  \\
	HC$_3$N			&96.6$\pm$0.1&5.3$\pm$0.1& 5.8(14)			& 190		&	-		&	&96.3$\pm$0.2&4.4$\pm$0.5	& - 		& -		& \\
	DCN				&96.8$\pm$0.1&5.8$\pm$0.3	& 3.6(14)			& 190		&	-		&	&95.6$\pm$0.1&1.7$\pm$0.3	& - 		& -		&  \\
	CO				& - 			& -			& >3.0(18) *		& 190	&	-		&	&  -			&	-		& -			& -		& \\
	\hdashline
	CH$_3^{18}$OH	&97.3$\pm$0.7&4.8$\pm$1.1& 1.6\±0.8(16)		& 170\±60	&	-		&	& \multicolumn{5}{c}{\multirow{8}{*}{Not detected}} \\
	$^{13}$CH$_3$CN	&97.1$\pm$0.3&5.1$\pm$0.7& 5.7\±1.3(14)		& 130\±60	&	-		&	& \multicolumn{5}{c}{} \\
	HC(O)NH$_2$		&97.3$\pm$0.4&6.2$\pm$0.9& 1.7\±0.7(15)		& 260\±120	&	-		&	& \multicolumn{5}{c}{} \\
	C$_2$H$_5$OH	&97.3$\pm$0.4&5.0$\pm$1.0& 2.4\±0.5(16)		& 85\±20		&	-		&	& \multicolumn{5}{c}{} \\
	H$_2$C$^{34}$S	&96.8$\pm$0.3&4.9$\pm$0.6& 1.4(15)			& 190		&	-		&	& \multicolumn{5}{c}{} \\
	CH$_3$COCH$_3$	&97.4$\pm$0.5&4.7$\pm$0.6& 2.0(16)			& 190		&	-		&	& \multicolumn{5}{c}{} \\
	C$_2$H$_3$CN	&96.2$\pm$0.3&7.2$\pm$0.7	& 2.9(15) 			& 190		&	-		&	& \multicolumn{5}{c}{} \\
	C$_2$H$_5$CN	&97.2$\pm$0.1&4.9$\pm$0.4& 4.5\±1.5(15)		& 170\±70	&	-		&	& \multicolumn{5}{c}{} \\
	\hline
\end{tabular}

\medskip
\textit{Notes}: $a(b)=a\times10^b$.  {*} Only a lower limit of the column density can be derived for the molecules where the emission is self-absorbed or filtered by the interferometer.   
\label{table:MoleculeProp}
\end{table*}

   \subsection{Critical density and opacity}
   \label{subsec:isotop}
 
The critical densities are estimated as $n_{\rm crit} = A_{\rm ij}/q_{\rm ij}$ with  $A_{\rm ij}$ the Einstein coefficient and $q_{\rm ij}$ the collision coefficient rates. The values of the critical densities for the individual transitions of the molecules, available in the Leiden Atomic and Molecular Database\footnote{\url{https://home.strw.leidenuniv.nl/~moldata/}.} \citep{schoier05}, are indicated in Tables \ref{table:LinesC3} and \ref{table:LinesC6}. We find values between $\sim$10$^4$ and $10^7$ cm$^{-3}$ for our lines, which are lower than the obtained $n_{\rm H_2}$ values of the cores ($\sim$10$^{9}$ cm$^{-3}$, see Table \ref{table:sources}). In the following we will hence assume that LTE conditions are verified.
 
 In general, isotopologues are less likely to be optically thick than their main isotopologue because of their lower abundances. Considering a distance of W43-MM1 from the Galactic centre of 4.5\,kpc \citep{zhang14}, we take for the isotopic ratios: $^{12}$C/$^{13}$C = 45$\pm$13 \citep{milam05}, $^{32}$S/$^{33}$S = 90$\pm$30 \citep{chin96}, and $^{18}$O/$^{16}$O = 327$\pm$32 \citep{wilson94}. The detection of $^{13}$CH$_3$OH in core \#6 is important since the methanol is optically thick. We also preferred to use the available isotopologues of CS and OCS to avoid some underestimations of abundances.

   \subsection{Temperatures and abundances}
                The total column density $N_{\rm tot}$ and the excitation temperature $T_{\rm ex}$ can be obtained in LTE and optically thin conditions using the relation 
        \begin{equation}
                ln\frac{N_{\rm u}}{g_{\rm u}} = lnN_{\rm {tot}}-lnQ(T_{\rm ex})-\frac{E_{\rm u}}{k_{\rm B}T_{\rm ex}}
        ,\end{equation}
   
   with $N_{\rm u}$ the upper level column density, $g_{\rm u}$ the degeneracy of the upper state, $Q(T_{\rm ex})$ the partition function, and $E_{\rm u}$ the upper state energy. We assume here that the filling factor is equal to 1. In LTE conditions, the excitation temperature is identical to the kinetic temperature.
   
For core \#3, we could derive the column density and the excitation temperature for ten molecules out of the 25 molecules detected using rotational diagrams since we detected more than two transitions for these ten molecules. For core \#6, it was only possible for methanol, acetaldehyde, and methyl formate, which have enough detected transitions to build a rotational diagram. The rotational diagrams were performed using the CASSIS\footnote{\url{http://cassis.irap.omp.eu}.} software \citep{vastel15}. The derived values are displayed in Table \ref{table:MoleculeProp}. Despite the significant uncertainties, the temperature seems to be lower for core \#6 (15-90\,K) than for core \#3 ($\sim $190\,K). 

The local density is high enough to reach LTE conditions, so that we could use the equation \citep{tielens05}
        \begin{equation}
                N_{\rm tot} = \frac{1.94\times10^3}{A_{\rm ij}} \nu_{\rm ij}^2 \int{T_{\rm mb} d\nu}\frac{Q(T_{\rm ex})}{g_{\rm u}\exp(-E_{\rm u}/k_{\rm B} T_{\rm ex})} ,
        \label{eq:relationNT}
        \end{equation}

        where $N_{\rm tot}$ is in cm$^{-2}$, $\nu_{\rm ij}$ is the rest frequency in GHz, and the only parameter coming from the observations is the integrated intensity $\int{T_{\rm mb} d\nu}$ in K.km.s$^{-1}$. The spectroscopic parameters come from the Cologne Database for Molecular Spectroscopy (CDMS\footnote{http://www.astro.uni-koeln.de/cdms/.}) and the Jet Propulsion Laboratory (JPL\footnote{http://spec.jpl.nasa.gov/.}) catalogue depending on the molecules (see Tables \ref{table:LinesC3} and \ref{table:LinesC6}). When it was not possible to use a rotational diagram, we assumed an excitation temperature of 190K for core \#3 to derive a column density with  Eq. (\ref{eq:relationNT}). These results are the inputs for the synthetic spectra in Figs. \ref{fig:Spectra-core3} and \ref{fig:Spectra-core6}. Residuals show that only a few lines are unidentified towards core \#3. Since H$_2$CO and C$^{18}$O are self-absorbed and SiO and CO are filtered, only lower limits can be obtained and no correct synthetic spectra can be performed for these molecules. For the optically thick molecules, a correction of the column density is made with their observed isotopologues; the corrected values are in Table \ref{table:Nisotop} for core \#3 and in Table \ref{table:Ndiffscenario} for core \#6. The O$^{13}$CS/OC$^{33}$S ratio for core \#6 and the O$^{13}$CS/OC$^{33}$S and $^{13}$CH$_3$OH/CH$_3^{18}$OH ratios for core \#3 are consistent with the elemental isotopic ratios mentioned in Sect. \ref{subsec:isotop}. 
        
\begin{table*}                    
\begin{center}
\setlength{\tabcolsep}{3.5pt}
\caption[]{List of sources used for comparison with cores \#3 and \#6.}

\begin{tabular}{lcccccccc} 
        \hline\hline
        Source                  & Name                          & Nature of              & Hot core/     & Distance              & Size          & Mass            & $L_{\rm bol}$                 & Reference  \\
                                        & in Fig. \ref{fig:results}     & the core                & corino                & [kpc]                         & [au]            &                       & $[L_\odot]$           & \\
        \hline
        L134N                   & L134N                 & Prestellar            & -                       & 0.2                   & 1500          & LM            & $10^{-2}$               & \cite{ohishi92}                       \\
        B1b-S                   & B1b-S                 & Young protostar         & Hot corino    & 0.2                   & 80                    & LM              & $10^{-1}$             & \cite{marcelino18}                    \\      
        B1b-N                   & B1b-N                 & FHSC candidate        & -                       & 0.2                   & 500           & LM            & $10^{-1}$               & \cite{gerin15}                \\ 
        L1544                   & L1544                 & Prestellar            & -                       & 0.1                   & >4000         & LM            & $10^0$                  & \cite{vastel14, vastel18}                     \\
        L1689B                  & L1689B                        & Prestellar            & -                       & 0.1                   & 3500          & LM            & $10^0$                  & \cite{bacmann12}  \\  
        NGC1333 IRAS4 A2& IRAS4A2               & Protostellar          & Hot corino      & 0.2                   & 250           & LM            & $10^1$                  & \cite{lopezsepulcre17}  \\ 
        IRAS 16293-2442 & IRAS16293             & Protostellar          & Hot corino      & 0.1                   & 150           & LM            & $10^1$                  & \cite{cazaux03}       \\ 
        W43-MM1 \#6             & Core 6                        & Prestellar ?               & ?                     & 5.5                   & 1300           & HM            & $10^1$                        & \cite{nony18} \\ 
        CepE-A                  & CepE-A                        & Protostellar          & Hot corino      & 0.7                   & 510           & IM                    & $10^2$                  & \cite{ospina18}  \\
        NGC 7129 FIRS2  & FIRS2                         & Protostellar          & Hot core        & 1.2                   & 1900          & IM                    & $10^2$                  & \cite{fuente14}               \\
        NGC 2264 CMM3A  & CMM3A         & Protostellar          & Hot core      & 0.7                     &  200          & IM                    & $10^2$                        & \cite{watanabe17}       \\ 
        NGC 7538S MM3   & MM3                   & Prestellar ?          & ?                       &                               & 1100          & IM                      &                               &                       \\      
        \hspace{4.9em} MM2& MM2                 & Protostellar          & Hot core        & 2.6                   & 1100          & IM                    & $10^4$  $^ a$& \cite{feng16}                    \\      
        \hspace{4.9em} MM1& MM1                 & Protostellar          & Hot core        &                               & 1100          & HM            &                               &                         \\      
        CygX-N63 hot core       & N63hc         & \multirow{2}{*}{Young protostar}  & \multirow{2}{*}{Hot core}    & \multirow{2}{*}{1.4}  & 500           & HM      & \multirow{2}{*}{$10^2$ $^ a$}         & \multirow{2}{*}{\cite{fechtenbaumthesis}}         \\
        \hspace{1.24em}lukewarm region& N63lw   &                       &                                 &                       & 2100          & HM      &                                       &               \\
        W43-MM1 \#3             & Core 3                        & Protostellar                 & Hot core      & 5.5                   & 1200          & HM              & $10^3$                        & \cite{nony18} \\ 
        G35.20-0.74N A  & G35.20                        & Protostellar          & Hot core        & 2.2                   & 1300          & HM            & $10^4$                  & \cite{allen17}                        \\
        G35.03+0.35 A   & G35.03                        & Protostellar          & Hot core        & 2.3                   & 1100          & HM            & $10^4$                  & \cite{allen17}                        \\
        G34.26+015              & G34.26                        & Protostellar          & Hot core        & 3.7                   & 3500          & HM            & $10^4$                  & \cite{mookerjea07}\\
        \multirow{2}{*}{Orion KL}       & \multirow{2}{*}{KL}           & \multirow{2}{*}{Protostellar}   & \multirow{2}{*}{Hot core} & \multirow{2}{*}{0.4} & \multirow{2}{*}{4000} & \multirow{2}{*}{HM} & \multirow{2}{*}{$10^5$}                 & \cite{sutton95}\\
                                        &                       &                               &                       &                               &                       &                       &                                       & \cite{crockett14} \\ 
        Sgr B2(N2)              & SgrB2                 &  Protostellar         & Hot core        & 8.3                   & 12000         & HM            & $10^6$                  & \cite{belloche16}     \\
        \hline
\end{tabular}
\end{center}
\textit{Notes}: 
The distance is from the Sun. LM, IM, and HM stand for low-mass ($M<2M_\odot$), intermediate-mass ($2M_\odot<M<8M_\odot$), and high-mass ($M>8M_\odot$), respectively. The third column indicates if the protostellar cores harbour a hot core or a hot corino.

$^ a$ Total value from all the sources mentioned.
\label{table:SourcesLit}
\end{table*}

        The obtained abundances relative to H$_2$ are displayed in Fig \ref{fig:results}. We consider that the depth of the molecular distribution along the line of sight is equal to the size of the continuum cores, hence we use column density values $N_{\rm H2} =  (2.6\pm0.1) \times 10^{25}{\rm\,cm}^{-2}$ and $N_{\rm H2}= (1.9\pm0.3) \times 10^{25}{\rm\,cm}^{-2}$ (for core \#3 and \#6, respectively) from $n_{\rm H_2}$ densities in Table \ref{table:sources} for a source size equal to the size of the beam. Only the molecules centred on the cores are selected because they are more likely to be directly associated to the core. We chose to take all the abundances derived at the mean temperature $T_{\rm ex}=190$ K for core \#3. To investigate the nature of core \#6, we have derived the abundances in the typical range of temperature for star-forming regions ($T_{\rm ex}=10-200$ K). The values observed for various star-forming regions (see Table 5) are plotted for comparison and the nature of the cores are discussed in the next section. 
        
\begin{table}                    
        \begin{center}
                \setlength\tabcolsep{5mm}
                \caption[]{Column densities in core \#3 estimated from isotopologues.}

                \begin{tabular}{lc} 
                \hline\hline
                Molecule                  & $N$[cm$^{-2}$] \\ 
                \hline
                CH$_3$OH        &       5.1\±2.7(18)            \\
                CS                      &       4.0\±1.2(16)            \\
                OCS                     &       1.4\±0.5(17)            \\
                H$_2$CO         &       9.4\±2.7(16)            \\
                CO                      &       >5.3(19)                        \\
                \hline
                \label{table:Nisotop}
                \end{tabular}
        \end{center}
\textit{Notes}: $a(b)=a\times10^b$. The methanol density is derived from the rotational diagrams of its isopotologues. For the other molecules, the densities are obtained using isotopologue column densities at 190\,K. The source size is assumed to be equal to the beam.
\end{table}

        \begin{figure*} \centering
                \includegraphics[width=1.0\linewidth]{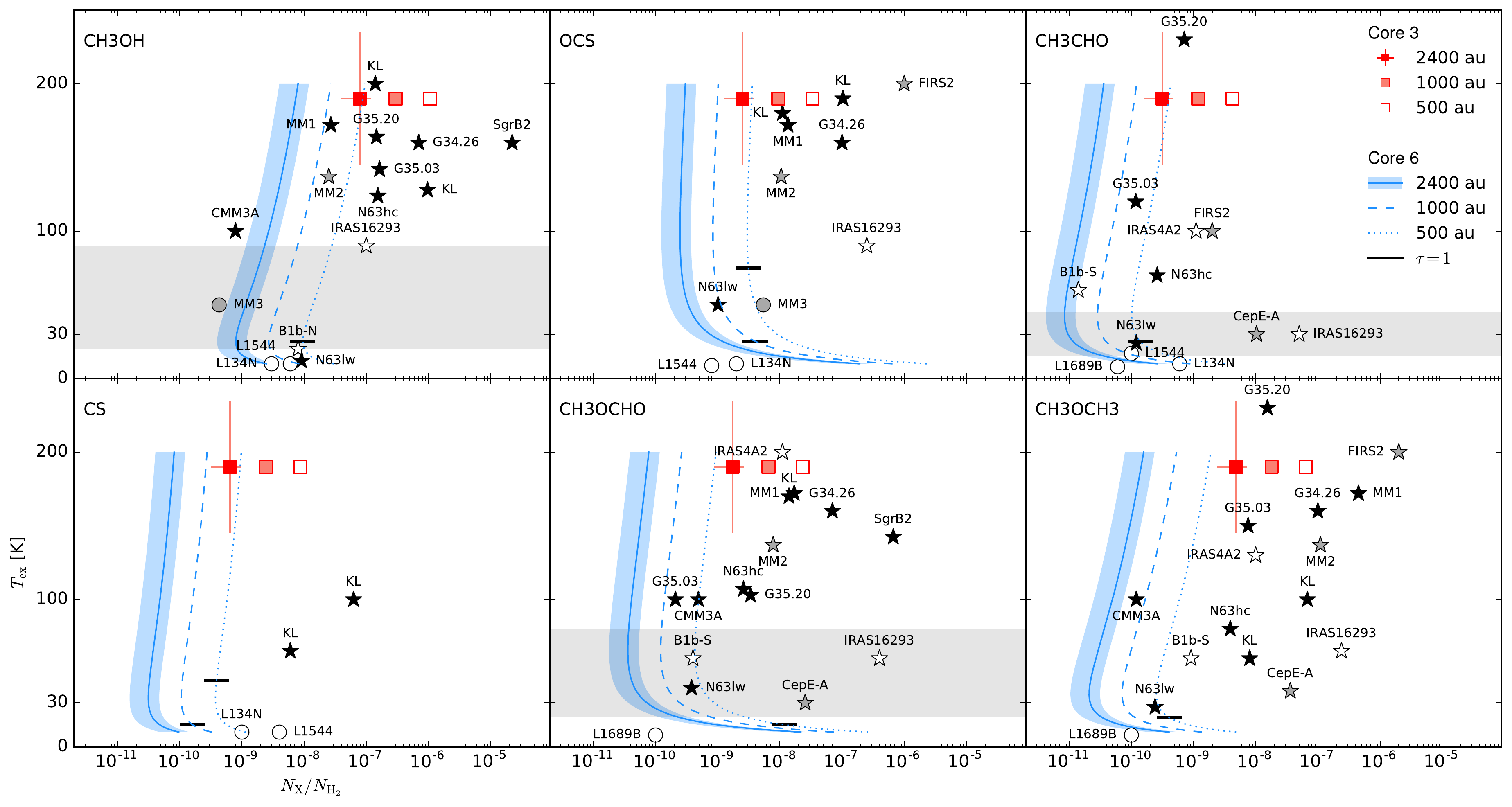}
                \caption {Abundances with respect to H$_2$ for selected molecules. Results for core \#3 (red squares) and core \#6 (blue lines) considering different source sizes (see legend) are from LTE modelling. The temperature $T_{\rm ex}=190$ K is fixed for core \#3, it varies from 10 to 200 K for core \#6. The blue area represents an error of 50\% on the estimation of the abundance, which includes the error on the line area and on the value of $N_{\rm H_2}$. The grey area is the temperature range estimated for core \#6 from the rotational diagrams. For a given size, a molecule becomes optically thick below the black line. The sources listed in Table 5 are represented by a dot for prestellar cores or a star for protostellar cores. The colour code is white for low-mass, grey for intermediate-mass, and black for high-mass cores. The names refer to column 2 of Table 4.}
                \label{fig:results}
        \end{figure*}
        
\section{Discussion}

        When considering the distribution of the molecules around core \#6, the high density component observed at the north-east of core \#6 may be associated to the heating of the outer envelope of the core by the interaction with a shock. Core \#6 appears to be in a large-scale shock, traced by the SiO distribution in Fig. \ref{fig:Mdecentred}, and \cite{louvet16} showed that the narrow component of the SiO emission comes from the low-velocity shock spread along the length of the ridge. We note also that \cite{nony18} showed that an outflow seen in SiO and CO emission comes from core \#18 in this direction.
        
         The lines detected towards core \#6 are narrower than towards core \#3, with a mean line width of 3.2\±0.1 km/s against 5.2\±0.2 km/s. Narrower lines for core \#6 are a first clue to propose that it is less active and thus less evolved than its protostellar neighbour. The difference in state of evolution cannot be due to the W-R/OB stars association of W43, which is located $\sim 5$ pc away, as the cores are out of the ionisation front caused by their UV radiation \citep{motte03, nguyen17}. Furthermore the cores do not show any inner velocity gradient at the spatial and spectral resolutions of our observations.
         
         The COMs detected in core \#6 are usually detected in hot cores, where they are formed on grain mantles and then desorbed during the warm-up phase. However, the detection of these molecules in low-mass prestellar cores \citep{bacmann12, cernicharo12,vastel14} shows that their presence is not sufficient to determine the nature of the core. Also, lower abundances of COMs in this core with respect to core \#3 is not a direct proof of a prestellar signature since other sources, like warm carbon chain chemistry sources, are poor in COMs \citep{sakai08}.
        
        Hereafter we compare the abundances observed  in core \#3 and \#6 to those in other sources considering two scenarios. In the first one, the size of the region emitting the detected molecules is taken as equal to the beam; in the second one, we consider that they are unresolved. The abundances taken in the literature for the other sources correspond to different source sizes and are not corrected by the filling factor or the H$_2$ column density. Nonetheless the trends remain the same: the molecules are more abundant in the high-mass hot cores than in hot corinos and even more than in low-mass prestellar cores. Thus it does not impact the qualitative discussion that follows.
                
\subsection{Cores sizes similar to the beam}
In this case, we consider no beam dilution. The observed abundances towards core \#3 are represented in Fig. \ref{fig:results} by red filled squares and those towards core \#6 by blue solid lines. First of all, the observed abundances are globally lower than for the other sources. Indeed we find column densities of the same order as for other sources but the local densities ($\sim 7\times10^9{\rm cm}^{-3}$) are huge compared to the usual $n_{\rm H_2} \sim 10^{5}-10^{8} {\rm cm}^{-3}$ observed in star-forming regions. Core \#3 abundances are at least an order of magnitude below for OCS and CH$_3$OCHO compared to the other high-mass protostellar cores. However, for CH$_3$OH and CH$_3$OCH$_3$, the core \#3 results are more consistent with some hot cores (G35.20-0.74N A and G35.03+0.35 A). All abundances derived for core \#6 are much lower than other sources at hot core and hot corino temperatures, except the CH$_3$OH value compatible with NGC 2264 CMM3A. 

The only sources comparable are cold low-mass prestellar cores (L1544, L1689B) and the very young low-mass protostar B1b-N, suggesting a temperature of about 20 K. The envelope on a 2400 au scale would be at the same temperature as the dust, whose temperature is around 23 K. This means that there is no internal source of heating (in other terms a protostar), since such a source would unbalance the two temperatures. At this low temperature, bright emission of the J=3-2 line of N$_2$D+ ($\nu =231.32183$ GHz, $E_{\rm up} =  22.2$ K) could be expected \citep{tobin19} but it is not detected, suggesting that the core is hotter than estimated in this scenario.
        
\subsection{Unresolved cores}
We suppose now that the size of the molecular emission is smaller than the beam size. The molecular column densities are corrected by the filling factor assuming a $r^{-2}$ density model for $n_{\rm H_2}$. When considering core sizes smaller than 500 au, all lines become optically thick in our temperature range and their profiles no longer match with our observations. With a core size of 500 au, $n_{\rm H_2} = 1.5\times10^{10} {\rm cm}^{-3}$ ($N_{\rm H_2} = 9.0\times10^{24} {\rm cm}^{-2}$) and $n_{\rm H_2} = 1.3\times10^{10} {\rm cm}^{-3}$ ($N_{\rm H_2} = 7.0\times10^{24} {\rm cm}^{-2}$) for cores \#3 and \#6, respectively. The abundances are thus enhanced by a factor of 12 with a source size of 500 au compared to a source size of 2400 au. It is also a way to mimic results from a source with the size of the beam but with a local density of $\sim 10^{8} {\rm cm}^{-3}$ compatible with young protostellar cores.

With a size of 500 au, results from core \#3 are more consistent with high-mass protostellar cores for OCS and CH$_3$OCHO (see red empty squares on Fig. \ref{fig:results}). However, the brightest lines become optically thick. Considering a size of 1000 au for core \#3 (pale red filled squares on Fig. \ref{fig:results}) seems a better compromise to match with the observed line profiles and the other core abundances at the same time. The corresponding column densities are listed in Table \ref{table:Ndiffscenario3}.

For core \#6, with a size of 500 au (blue dotted lines on Fig. \ref{fig:results}) the results are comparable to cores at various evolutionary stages between 20 and 90 K. However, some of the main lines are optically thick (e.g. $^{13}$CS and OCS) for temperatures lower than 75 K excluding the scenario of a compact cold core. This leaves the possibility of a compact core at a high temperature of 80 K, included in the temperature range given by the rotational diagrams and consistent with sources like the hot corino B1b-S and the protostellar core NGC 2264 CMM3A. By adjusting the size of core \#6 to 1000 au (blue dashed lines on Fig. \ref{fig:results}), most of the lines become optically thin and low temperatures are possible. In particular, the CH$_3$OH and CH$_3$CHO abundances match with L134N values at 20 K with a comparable size. The observed abundances for core \#6 with sizes 500-1000 au are also comparable to the lukewarm region of the young high-mass protostellar core CygX-N63 \citep[2100 au, ][]{fechtenbaumthesis}, but still at least an order of magnitude below the 500 au N63 hot core values.

NGC 7538 S MM1 is a high-mass protostellar object, more evolved than the hot molecular core NGC 7538 S MM2 and NGC 7538 S MM3, which is comparable to prestellar objects because it only exhibits emission of lower excitation lines. While the core \#3 results match with NGC 7538 S MM1, core \#6 is more likely to be similar to NGC 7538 S MM3. Also, B1b-S displays a spectrum as rich in lines as core \#3, and the B1b-N molecular emission shown in \cite{marcelino18} is as poor as core \#6. However, the CH$_3$OH emission in B1b-N is associated to a nascent outflow of size 1300 au \citep{gerin15}, unresolved at our beam size, but with abundances comparable to core \#6. 

\begin{table}[htb]                  
        \begin{center}
                \setlength\tabcolsep{3mm}
                \renewcommand{\arraystretch}{1.02}
                \caption[]{Column densities in core \#6 estimated for different scenarios.}
                \begin{tabular}{lcccc} 
                \hline\hline
                Size                            & 2400 au       & 1000 au         & 1000 au       & 500 au\\
                 $T_{\rm ex}$           & 20 K          & 30 K          & 80 K            & 80 K  \\
                \hline 
                Molecule        & \multicolumn{4}{c}{$N$ [cm$^{-2}$]}\\
                \hline  
                C$^{18}$O               & 1.3(16)               & 4.9(16)               & 9.4(16)         & 3.4(17)               \\      
                CH$_3$OH                & 2.9(17)               & 1.4(17)               & 3.0(16)         & 1.1(17)               \\
                $^{13}$CH$_3$OH & 1.4(15)               & 4.6(15)               & 1.1(16)         & 4.0(16)               \\      
                CH$_3$CHO               & 9.4(14)               & 1.8(15)               & 2.1(15)         & 7.6(15)               \\      
                $^{13}$CS               & 4.5(13)               & 1.3(14)               & 1.7(14)         & 6.2(14)               \\      
                CH$_3$OCH$_3$   & 3.8(15)               & 8.3(15)               & 1.4(16)         & 5.0(16)               \\      
                CH$_3$OCHO              & 1.3(16)               & 1.4(16)               & 7.0(15)         & 2.5(16)               \\      
                OCS                             & 2.9(16)               & 2.3(16)         & 6.1(15)               & 2.2(16)               \\      
                O$^{13}$CS              & 3.2(15)               & 3.1(15)               & 1.0(15)         & 3.7(15)\\     
                OC$^{33}$S              & 1.8(15)               &1.7(15)                & 5.9(14)         & 2.1(15)\\     
                \hline
                \multicolumn{5}{l}{Column densities estimated from the isotopologues}\\
                CH$_3$OH                & 6.3(16)               & 2.1(17)               & 5.0(17)         & 1.8(18) \\
                CS                              & 1.9(15)               & 5.9(15)         & 7.7(15)               & 2.8(16) \\
                OCS                             & 1.5(17)               & 1.5(17)         & 4.9(16)               & 1.8(17) \\
                CO                              & 4.3(18)               & 1.6(19)         & 3.1(19)               & 1.1(20)\\
                \hline
                \label{table:Ndiffscenario}
                \end{tabular}
        \end{center}
\textit{Note}: $a(b)=a\times10^b$.
\end{table}

\section{Conclusion}

The study of the molecular content of core \#6 is challenging because of the few and weakly observed lines. We have identified 17 molecules in core \#6 based on the analysis of its line-rich neighbour core \#3. The distribution of these molecules reveals that ten are directly associated to core \#6 and the others can be associated to the interaction between its envelope and the low-velocity shock of the ridge.
The determination of the column densities and the temperatures from rotational diagrams was only possible for three molecules, methanol (CH$_3$OH), methyl formate (CH$_3$OCHO), and acetaldehyde (CH$_3$CHO). Because of the low signal-to-noise ratio and the range of upper energy levels, there is a huge uncertainty on the temperature (15-90 \,K) obtained by this method. To better constrain the temperature and core nature, we have compared our observations to the abundances of different sources including prestellar cores, hot corinos, and hot cores. The conclusion about the nature of core \#6 clearly depends on the estimated size of the core : 
\begin{itemize}
\item at the size of the beam, 2400 au, the core is cold ($T_{\rm ex} \approx T_{\rm dust}$) and may be prestellar or in a transition phase to a protostellar core; however, the non detection of the N$_2$D+ emission suggests that the temperature is greater than 20 K.
\item if the core size is 1000 au, it could be at the very beginning of the protostellar phase at 20-30 K or a young protostellar core at $\sim$80 K not active enough to display large outflows;
\item considering a size of 500 au, the core still cannot be comparable to hot cores but only to a hot corino with a temperature of $\sim$80 K.
\item the source size is not less than 500 au because the lines would be optically thick, which is not observed.
\end{itemize}
Core \#6 is not as evolved as core \#3, the absence of outflows in CO and SiO suggests that it does not contain a protostar. From the molecular study it appears as a good high-mass prestellar core candidate and deep dedicated line surveys with a higher spatial resolution are still necessary to investigate whether it harbours a hot corino, like those found towards low-mass class 0 protostars.

\begin{acknowledgements}
        This paper makes use of the following ALMA data: ADS/JAO.ALMA\#2013.1.01365.S, ADS/JAO.ALMA\#2015.1.01273.S. ALMA is a partnership of ESO (representing its member states), NSF (USA), and NINS (Japan), together with NRC (Canada), MOST and ASIAA (Taiwan), and KASI (Republic of Korea), in cooperation with the Republic of Chile. The Joint ALMA Observatory is operated by ESO, AUI/NRAO, and NAOJ. 
        This work was supported by the Programme National de Physique Stellaire and Physique et Chimie du Milieu Interstellaire (PNPS and PCMI) of CNRS/INSU (with INC/INP/IN2P3).
\end{acknowledgements}

%
%
\bibliographystyle{aa}  
\bibliography{aa35497-19}

\newpage
\clearpage
\onecolumn

\newpage
\begin{table}[p]                        
\setlength\tabcolsep{1mm}
\renewcommand\thetable{A\arabic{table}} 
\setcounter{table}{0}                   

\caption{Spectroscopic parameters of the transitions detected towards core \#3 and observed line parameters.}
\setlength{\tabcolsep}{5pt}
\renewcommand{\arraystretch}{0.95}
\begin{tabular}{cccccccc|ccc}
        \hline\hline
        Frequency       & Molecule              & Transition                            & $E_{\rm up}$    &S$\mu^2$& log($A_{\rm ij}$)    & $n_{\rm crit}$        &Database       & Line width       & Velocity              & $\int{T_{\rm mb} d\varv}$      \\ 
        $$[MHz]         &                               & J=                                            & [K]             & [D$^2$]       & [s$^{-1}$]    & [cm$^{-3}$]   &                       & [km.s$^{-1}$]           & [km.s$^{-1}$]                 & [K.km.s$^{-1}$] \\
        \hline
        216109.78       & CH$_3$OCHO    & $19_{2,18}-18_{2,17}$ E       & 109.3   & 49.4  & -3.83         & -                     & JPL           & 3.9$\pm$0.1             & 97.2$\pm$0.1          & 33.6$\pm$1.2 \\ 
        216115.57       & CH$_3$OCHO    & $19_{2,18}-18_{2,17}$ A       & 109.3   & 49.4  & -3.83         & -                     & JPL           & 4.9$\pm$0.1             & 97.3$\pm$0.1          & 41.3$\pm$1.4 \\ 
        216147.36       & OC$^{33}$S            & $18-17$                               & 98.6    & 9.2   & -4.04                 & 1.3(6) $^a$   & CDMS          & 3.7$\pm$0.2     & 96.6$\pm$0.1     & 8.0$\pm$0.8 \\ 
        216210.91       & CH$_3$OCHO    & $19_{1,18}-18_{1,17}$ E       & 109.3   & 49.4  & -3.83         & -                     & JPL           & 4.4$\pm$0.1             & 97.2$\pm$0.1          & 37.4$\pm$1.4 \\ 
        216216.54       & CH$_3$OCHO    & $19_{1,18}-18_{1,17}$ A       & 109.3   & 49.4  & -3.83         & -                     & JPL           & 5.1$\pm$0.2             & 97.4$\pm$0.1          & 40.4$\pm$2.8 \\ 
        \hdashline
        217044.62       & $^{13}$CH$_3$OH& $14_{1,13}-13_{2,12}$ - -    & 254.3   & 5.8   & -4.62         & 1.2(7) $^a$   & CDMS          & 3.1$\pm$0.2           & 97.6$\pm$0.1            & 15.8$\pm$2.1 \\ 
        217104.98       & SiO v=0               & $5-4$                                 & 31.3    & 48.0  & -3.28         & 4.8(6)                & CDMS          & 65$^b$                  & 91.3$^b$                      & >86 $^b$\\ 
        217191.40       & CH$_3$OCH$_3$ & $22_{4,19}-22_{3,20}$         & 253.4   & 327.7 & -4.27         & -                     & CDMS          & 6.5$\pm$0.2     & 96.9$\pm$0.1     & 42.9$\pm$2.0 \\ 
        217238.54       & DCN                   & $3-2$                                 & 20.9    & 80.5  & -3.34                 & -                     &CDMS           & 5.8$\pm$0.3     & 96.8$\pm$0.1     & 51.4$\pm$4.5 \\ 
        \hdashline
        218127.21       & CH$_3$COCH$_3$& $20_{2,18}-19_{3,17}$ EE      & 119.1   & 1200.4        & -3.66         & -                     & JPL           & 5.6$\pm$0.2     & 97.0$\pm$0.1     & 16.1$\pm$1.2 \\ 
        218199.00       & O$^{13}$CS            & $18-17$                               & 99.5    & 9.2   & -4.52         & 4.1(5) $^a$   & CDMS          & 4.0$\pm$0.1     & 97.1$\pm$0.1     & 19.1$\pm$0.9 \\ 
        218222.19       & H$_2$CO               & $3_{0,3}-2_{0,2}$                     & 21.0    & 16.3  & -3.55         & 2.6(6)                & CDMS          & 17$^b$                  & 98.0$^b$                      & >52 $^b$\\ 
        218280.90       & CH$_3$OCHO    & $17_{3,14}-16_{3,13}$ E       & 99.7    & 43.6  & -3.82         & -                     & JPL           & 4.2$\pm$0.1             & 97.2$\pm$0.1          & 34.5$\pm$1.1 \\ 
        218297.89       & CH$_3$OCHO    & $17_{3,14}-16_{3,13}$ A       & 99.7    & 43.6  & -3.82         & -                     & JPL           & 4.2$\pm$0.1             & 97.1$\pm$0.1          & 32.9$\pm$1.3 \\ 
        218324.72       & HC$_3$N               & $24-23$                               & 131.0   & 334.2 & -3.08         & 1.7(7)                & CDMS          & 5.3$\pm$0.1     & 96.6$\pm$0.1     & 73.0$\pm$2.0 \\ 
        \hdashline
        219505.59       & C$_2$H$_5$CN  & $24_{2,22}-23_{2,21}$         & 135.6   & 353.2 & -3.05         & -                     & JPL           & 4.7$\pm$0.1     & 96.9$\pm$0.1     & 36.0$\pm$1.3 \\ 
        219560.35       & C$^{18}$O             & $2-1$                                 & 15.8    & 0.02  & -6.22         & 9.9(3) $^a$   & CDMS          & 4$^b$                 & 96.0$^b$                        & >37 $^b$\\ 
        \hdashline
        219908.53       & H$_2$$^{13}$CO& $3_{1,2}-2_{1,1}$                     & 32.9    & 43.5  & -3.59         & 2.3(6) $^a$   & CDMS          & 4.6$\pm$0.2     & 97.2$\pm$0.1     & 35.7$\pm$2.5 \\
        219949.44       & SO                    & $6_{5}-5_{4}$                 & 35.0    & 14.0  & -3.87         & 2.3(6)                & CDMS          & 5.6$\pm$0.2     & 97.0$\pm$0.1     & 56.0$\pm$2.2 \\ 
        219983.68       & CH$_3$OH              & $25_3-24_4$ E1                        & 802.2   & 8.4   & -4.70         & -                     & JPL           & 4.4$\pm$0.2             & 97.3$\pm$0.1          & 18.4$\pm$1.5 \\
        219993.66       & CH$_3$OH              & $23_5-22_6$ E1                        & 775.9   & 6.6   & -4.76         & -                     & JPL           & 4.2$\pm$0.3             & 97.6$\pm$0.1          & 16.9$\pm$1.8 \\ 
        \hdashline
        230315.79       & CH$_3$CHO     & $12_{2,11}-11_{2,10}$ E       & 81.1    & 147.4 & -3.38         & -                     & JPL           & 5.8$\pm$0.3             & 95.6$\pm$0.1          & 36.5$\pm$2.6 \\ 
        230368.76       & CH$_3$OH              & $22_4-21_5$ E1                        & 682.8   & 6.6   & -4.68         & -                     & JPL           & 5.0$\pm$0.2             & 97.6$\pm$0.1          & 30.2$\pm$2.2 \\ 
        230487.94       & C$_2$H$_3$CN  & $24_1-23_1$                   & 141.2   & 1045.0        & -2.50         & -                     & CDMS          & 8.0$\pm$0.6     & 96.2$\pm$0.2     & 20.4$\pm$2.4 \\ 
        230538.00       & CO                    & $2-1$                                 & 16.6    & 0.02  & -6.16         & 1.1(4)                & CDMS          & 85$^b$                  & 89.5$^b$                      & >550 $^b$\\ 
        230672.55       & C$_2$H$_5$OH  & $13_{2,11}-12_{2,10}$         & 138.6   & 20.3  & -3.97         & -                     & JPL           & 4.5$\pm$0.3     & 97.1$\pm$0.1     & 12.4$\pm$1.8 \\
        230738.56       & C$_2$H$_3$CN  & $25_0-24_0$                   & 145.5   & 1089.8        & -2.53         & -                     & CDMS          & 4.9$\pm$0.4     & 96.2$\pm$0.3     & 16.9$\pm$4.3 \\ 
        \hdashline
        231060.99       & OCS                   & $19-18$                               & 110.9   & 9.7   & -4.45         & 4.9(5)                & CDMS          & 5.7$\pm$0.1     & 96.5$\pm$0.1     & 129.2$\pm$3.5 \\ 
        231220.68       & $^{13}$CS             & $5-4$                                 & 33.3    & 38.3  & -3.60         & 4.3(6) $^a$   & JPL           & 5.0$\pm$0.1           & 97.0$\pm$0.1            & 78.9$\pm$2.8 \\ 
        231269.90       & CH$_3$CHO     & $12_{6,7}-11_{6,6}$ E         & 153.4   & 113.9 & -3.48         & -                     & JPL           & 5.0$\pm$0.2     & 97.3$\pm$0.1     & 15.4$\pm$1.1 \\ 
        231281.11       & CH$_3$OH              & $10_2$-$ - 9_3$-                      & 165.3   & 2.68  & -4.73         & 2.5(7)                & JPL           & 5.4$\pm$0.1             & 97.2$\pm$0.1          & 74.5$\pm$2.1 \\ 
        231310.42       & C$_2$H$_5$CN  & $26_{1,25}-25_{1,24}$         & 153.4   & 383.1 & -2.98         & -                     & JPL           & 4.9$\pm$0.1     & 96.7$\pm$0.1     & 48.0$\pm$1.5 \\ 
        231467.50       & CH$_3$CHO     & $12_{4,8}-11_{4,7}$ E         & 108.4   & 135.0 & -3.41         & -                     & JPL           & 5.3$\pm$0.3             & 96.2$\pm$0.1          & 11.6$\pm$1.0 \\ 
        231506.29       & CH$_3$CHO     & $12_{4,9}-11_{4,8}$ E         & 108.2   & 135.0 & -3.41         & -                     & JPL           & 6.0$\pm$0.6             & 97.3$\pm$0.2          & 17.2$\pm$2.7 \\ 
        \hdashline
        231595.27       & CH$_3$CHO     & $12_{3,10}-11_{3,9}$ E        & 92.6    & 142.3 & -3.39         & -                     & JPL           & 7.0$\pm$1.0             & 95.7$\pm$0.3          & 9.8$\pm$2.3 \\ 
        231686.68       & CH$_3^{18}$OH         & $5_{-0,5}-4_{-0,4}$ E         & 46.2    & 16.2  & -4.27         & 4.5(5) $^a$   & CDMS          & 4.6$\pm$0.4     & 97.4$\pm$0.1     & 7.8$\pm$1.2 \\ 
        231724.16       & CH$_3$OCHO    & $18_{4,14}-17_{4,13}$ E       & 300.8   & 45.8  & -3.75         & -                     & JPL           & 4.0$\pm$0.3             & 97.3$\pm$0.1          & 7.6$\pm$1.0 \\ 
        231735.83       & CH$_3^{18}$OH         & $5_{-1,5}-4_{-1,4}$ E         & 39.0    & 15.6  & -4.29         & 5.8(5) $^a$   & CDMS          & 5.2$\pm$0.3     & 96.5$\pm$0.1     & 11.6$\pm$1.2 \\ 
        231748.72       & CH$_3$CHO     & $12_{3,10}-11_{3,9}$ E        & 92.5    & 141.1 & -3.39         & -                     & JPL           & 5.7$\pm$0.2             & 96.5$\pm$0.1          & 12.3$\pm$0.9 \\ 
        231758.45       & CH$_3^{18}$OH & $5_{0,5}-4_{0,4}$ A           & 33.4    & 16.2  & -4.27         & 4.5(5) $^a$   & CDMS          & 4.7$\pm$0.3     & 97.1$\pm$0.1     & 9.0$\pm$1.1 \\
        231796.52       & CH$_3^{18}$OH         & $5_{3,2}-4_{3,1}$ A           & 83.5    & 10.3  & -4.47         & 4.2(5) $^a$   & CDMS          & 5.4$\pm$1.4     & 97.8$\pm$0.4     & 8.6$\pm$3.4 \\ 
        231801.47       & CH$_3^{18}$OH & $5_{2,4}-4_{2,3}$ A           & 70.8    & 13.8  & -4.34         & 5.4(5) $^a$   & CDMS          & 4.2$\pm$0.5     & 96.8$\pm$0.1     & 8.0$\pm$1.4 \\ 
        231826.74       & CH$_3^{18}$OH & $5_{1,4}-4_{1,3}$ E           & 54.1    & 16.2  & -4.27         & 5.6(5) $^a$   & CDMS          & 4.5$\pm$0.5     & 97.1$\pm$0.2     & 6.7$\pm$1.2 \\ 
        231840.93       & CH$_3^{18}$OH & $5_{2,3}-4_{2,2}$ A           & 70.9    & 13.8  & -4.34         & 5.4(5) $^a$   & CDMS          & 4.2$\pm$0.7     & 97.4$\pm$0.2     & 6.3$\pm$1.5 \\ 
        231847.58       & CH$_3$CHO     & $12_{3,9}-11_{3,8}$ E         & 92.6    & 141.2 & -3.39         & -                     & JPL           & 6.2$\pm$0.8             & 97.1$\pm$0.2          & 13.3$\pm$2.7 \\ 
        231864.50       & CH$_3^{18}$OH & $5_{2,3}-4_{2,2}$ E           & 55.8    & 13.4  & -4.36         & 5.2(5) $^a$   & CDMS          & 4.0$\pm$0.2     & 97.3$\pm$0.1     & 6.0$\pm$0.7 \\ 
        231896.06       & CH$_3$OCHO     & $19_{4,16}-18_{4,15}$ E      & 309.7   & 48.14 & -3.75                 & -                     & JPL           & 4.5$\pm$0.2             & 97.2$\pm$0.1          & 8.5$\pm$0.7 \\ 
        231903.90       & CH$_3$OCHO    & $19_{12,7}-18_{12,6}$ A       &395.1  & 30.46   & -3.95         & -                     & JPL           & 6.4$\pm$0.5           & 97.5$\pm$0.2            & 8.2$\pm$1.1 \\ 
        231968.39       & CH$_3$CHO     & $12_{3,9}-11_{3,8}$ A         & 92.6    & 142.3 & -3.38         & -                     & JPL           & 4.8$\pm$0.3             & 96.9$\pm$0.1          & 12.1$\pm$1.3 \\ 
        231987.93       & CH$_3$OCH$_3$ & $13_{0,13}-12_{1,12}$         & 80.9    & 271.8 & -4.04         & -                     & CDMS          & 6.8$\pm$0.3     & 96.2$\pm$0.1     & 57.1$\pm$4.7 \\ 
        232034.63       & C$_2$H$_5$OH  & $18_{5,14}-18_{4,15}$         & 175.3   & 18.6  & -4.13         & -                     & JPL           & 5.4$\pm$0.5     & 96.8$\pm$0.2     & 6.3$\pm$1.2 \\ 
        232160.19       & CH$_3$OCHO    & $19_{9,10}-18_{9,9}$ E        & 353.3   & 39.3  & -3.83         & -                     & JPL           & 4.7$\pm$0.9     & 97.2$\pm$0.2     & 9.6$\pm$2.8 \\ 
        232164.44       & CH$_3$OCHO    & $19_{10,10}-18_{10,9}$ A      & 365.7   & 36.6  & -3.86         & -                     & JPL           & 4.6$\pm$0.7     & 97.5$\pm$0.3     & 12.7$\pm$3.0 \\ 
        232194.91       & $^{13}$CH$_3$CN& $13_3-12_3$                  & 142.4   & 757.2 & -2.99         & 5.2(6) $^a$   & JPL           & 5.1$\pm$0.3     & 97.1$\pm$0.1     & 17.4$\pm$2.1 \\ 
        232216.73       &  $^{13}$CH$_3$CN& $13_2-12_2$                 & 106.7   & 390.4 & -2.98         & 5.2(6) $^a$   & JPL           & 4.9$\pm$0.3     & 97.2$\pm$0.2     & 14.6$\pm$1.9 \\ 
        232229.82       & $^{13}$CH$_3$CN& $13_1-12_1$                  & 85.2    & 397.6 & -2.97         & 5.2(6) $^a$   & JPL           & 4.2$\pm$0.5     & 97.2$\pm$0.2     & 13.9$\pm$2.5 \\
        232234.19       & $^{13}$CH$_3$CN& $13_0-12_0$                  & 78.0    & 399.9 & -2.97         & 5.2(6) $^a$   & JPL           & 5.2$\pm$0.5     & 97.1$\pm$0.2     & 17.1$\pm$2.8 \\ 
        232273.65       & HC(O)NH$_2$   & $11_{2,10}-10_{2,9}$          & 78.9    & 139.0 & -3.05         & -                     & CDMS          & 7.3$\pm$0.2     & 97.5$\pm$0.1     & 13.2$\pm$1.1 \\ 
        232404.81       & C$_2$H$_5$OH  & $17_{5,13}-17_{4,14}$         & 160.1   & 17.5  & -4.14         & -                     & JPL           & 4.5$\pm$0.5     & 97.0$\pm$0.2     & 6.7$\pm$1.2 \\ 
        232418.52       & CH$_3$OH              & $10_2$+$-9_3$+                        & 165.4   & 2.68  & -4.73         & 2.2(7)                & JPL           & 5.4$\pm$0.4             & 97.1$\pm$0.1          & 38.8$\pm$3.5 \\ 
        \hline
\end{tabular}
\label{table:LinesC3}
\end{table}

\newpage
\begin{table}[p]                        
\setlength\tabcolsep{1mm}
\renewcommand\thetable{A\arabic{table}} 
\setcounter{table}{0}                   

\caption{continued.}
\renewcommand{\arraystretch}{0.95}
\setlength{\tabcolsep}{5pt}
\begin{tabular}{cccccccc|ccc}
        \hline\hline
        Frequency       & Molecule              & Transition                            & $E_{\rm up}$    &S$\mu^2$& log($A_{\rm ij}$)    & $n_{\rm crit}$        &Database       & Line width       & Velocity              & $\int{T_{\rm mb} d\varv}$      \\ 
        $$[MHz]         &                               & J=                                            & [K]             & [D$^2$]       & [s$^{-1}$]    & [cm$^{-3}$]   &                       & [km.s$^{-1}$]           & [km.s$^{-1}$]                 & [K.km.s$^{-1}$] \\
        \hline 
        232683.93       & CH$_3$OCHO    & $19_{10,10}-18_{10,9}$ E      & 365.5   & 36.7  & -3.86         & -                     & JPL           & 4.2$\pm$0.3     & 97.6$\pm$0.1     & 7.8$\pm$1.1 \\ 
        232738.62       & CH$_3$OCHO    & $19_{8,11}-18_{8,10}$ E       & 342.0   & 41.6  & -3.80         & -                     & JPL           & 4.5$\pm$0.2     & 97.2$\pm$0.1     & 11.0$\pm$0.8 \\ 
        232754.71       & H$_2$C$^{34}$S        & $7_{1,7}-6_{1,6}$                     & 57.9    & 55.9  & -3.74         & 2.0(6) $^a$   & CDMS          & 4.9$\pm$0.6     & 96.8$\pm$0.3     & 12.0$\pm$2.0 \\ 
        232783.45       & CH$_3$OH              & $18_3+ - 17_4+$                       & 446.54  & 5.46  & -4.66         & -                     & JPL           & 4.9$\pm$0.2             & 97.1$\pm$0.1          & 50.4$\pm$3.3 \\ 
        232790.02       & C$_2$H$_5$CN  & $26_{3,24}-25_{3,25}$         & 161.0   & 380.0 & -2.98         & -                     & JPL           & 5.6$\pm$0.4     & 97.0$\pm$0.1     & 46.5$\pm$5.1 \\ 
        232836.17       & CH$_3$OCHO    & $19_{8,12}-18_{8,11}$ A       & 341.8   & 41.6  & -3.81         & -                     & JPL           & 3.5$\pm$0.2     & 97.0$\pm$0.1     & 8.5$\pm$0.7 \\ 
        232839.68       & CH$_3$OCHO    & $19_{8,11}-18_{8,10}$ A       & 341.8   & 41.6  & -3.80         & -                     & JPL           & 4.2$\pm$0.7     & 97.5$\pm$0.3     & 9.5$\pm$2.2 \\ 
        232865.05       & CH$_3$COCH$_3$& $17_{8,10}-16_{7,9}$ EE       & 114.8   & 1121.9        & -3.53         & -                     & JPL           & 4.7$\pm$0.6     & 97.4$\pm$0.5     & 5.0$\pm$1.1 \\ 
        232945.80       & CH$_3$OH              & $10_{-3}-11_{-2}$ E2          & 190.4   & 3.0   & -4.67         & 1.6(7)                & JPL           & 5.8$\pm$0.3             & 97.0$\pm$0.1          & 78.7$\pm$6.3 \\
        232962.32       & C$_2$H$_5$CN  & $26_{10}-25_{10}$             & 261.6   & 328.4 & -3.04         & -                     & JPL           & 5.4$\pm$0.3     & 97.3$\pm$0.1     & 37.3$\pm$3.2 \\ 
        232967.57       & C$_2$H$_5$CN  & $26_{9}-25_{9}$                       & 240.9   & 339.2 & -3.03         & -                     & JPL           & 6.2$\pm$0.8     & 97.0$\pm$0.3     & 44.2$\pm$8.3 \\ 
        232975.51       & C$_2$H$_5$CN  & $26_{11}-25_{11}$             & 285.2   & 316.4 & -3.06         & -                     & JPL           & 5.3$\pm$0.1     & 96.8$\pm$0.1     & 34.4$\pm$1.2 \\ 
        232998.74       & C$_2$H$_5$CN  & $26_{8}-25_{8}$                       & 222.0   & 348.9 & -3.01         & -                     & JPL           & 4.6$\pm$0.5     & 97.3$\pm$0.2     & 36.7$\pm$5.9 \\ 
        233002.70       & C$_2$H$_5$CN  & $26_{12}-25_{12}$             & 310.7   & 303.3 & -3.07         & -                     & JPL           & 5.0$\pm$0.3     & 97.3$\pm$0.1     & 30.3$\pm$2.4 \\ 
        233041.09       & C$_2$H$_5$CN  & $26_{13}-25_{13}$             & 338.3   & 289.1 & -3.09         & -                     & JPL           & 5.1$\pm$0.3     & 97.3$\pm$0.1     & 22.8$\pm$2.4 \\ 
        233069.37       & C$_2$H$_5$CN  & $26_{7}-25_{7}$                       & 205.4   & 357.5 & -3.00         & -                     & JPL           & 5.0$\pm$0.2     & 96.9$\pm$0.1     & 42.6$\pm$3.3 \\
        233088.86       & C$_2$H$_5$CN  & $26_{14}-25_{14}$             & 368.2   & 273.6 & -3.12         & -                     & JPL           & 5.8$\pm$0.2     & 97.0$\pm$0.1     & 21.0$\pm$1.3 \\ 
        233193.36       & CH$_3$OCH$_3$ & $22_{2,21}-22_{1,22}$         & 232.9   & 135.6 & -4.56         & -                     & CDMS          & 4.6$\pm$0.6     & 96.8$\pm$0.2     & 17.1$\pm$3.7 \\ 
        233226.79       & CH$_3$OCHO    & $19_{4,16}-18_{4,15}$ A       & 123.2   & 48.0  & -3.74         & -                     & JPL           & 4.9$\pm$0.4     & 97.3$\pm$0.2     & 41.1$\pm$5.7 \\
        233246.79       & CH$_3$OCHO    & $19_{16,3}-18_{16,2}$ A       & 281.8   & 14.8  & -4.25         & -                     & JPL           & 4.6$\pm$0.4     & 96.5$\pm$0.2     & 11.0$\pm$1.6 \\ 
        233268.59       & CH$_3$OCHO    & $19_{16,4}-18_{16,3}$ E       & 281.8   & 14.8  & -4.25         & -                     & JPL           & 5.5$\pm$0.3     & 97.2$\pm$0.1     & 11.4$\pm$1.1 \\ 
        233277.94       & C$_2$H$_5$CN  & $26_{17}-25_{17}$             & 470.6   & 220.7 & -3.21         & -                     & JPL           & 4.5$\pm$0.6     & 97.3$\pm$0.2     & 7.7$\pm$2.0 \\ 
        233310.12       & CH$_3$OCHO    & $19_{15,5}-18_{15,4}$ A       & 261.3   & 19.1  & -4.14         & -                     & JPL           & 5.0$\pm$0.1     & 97.4$\pm$0.1     & 20.1$\pm$0.6 \\ 
        233315.78       & CH$_3$OCHO    & $19_{15,4}-18_{15,3}$ E       & 261.3   & 19.1  & -4.14         & -                     & JPL           & 5.8$\pm$0.6     & 97.0$\pm$0.2     & 13.2$\pm$2.1 \\ 
        233331.21       & CH$_3$OCHO    & $19_{15,5}-18_{15,4}$ E       & 261.3   & 19.1  & -4.14         & -                     & JPL           & 4.6$\pm$0.7     & 97.0$\pm$0.3     & 10.1$\pm$2.6 \\ 
        233394.66       & CH$_3$OCHO    & $19_{14,6}-18_{14,5}$ A       & 242.1   & 23.2  & -4.06         & -                     & JPL           & 6.0$\pm$0.3     & 98.5$\pm$0.1     & 34.9$\pm$3.1 \\ 
        233414.43       & CH$_3$OCHO    & $19_{14,6}-18_{14,5}$ E       & 242.1   & 23.2  & -4.06         & -                     & JPL           & 5.2$\pm$0.4     & 97.1$\pm$0.1     & 15.7$\pm$1.9 \\ 
        233443.10       & C$_2$H$_5$CN  & $26_{5,22}-25_{5,21}$         &178.8  & 371.1   & -2.98         & -                     & JPL           & 5.7$\pm$0.4     & 97.3$\pm$0.1     & 42.4$\pm$4.8 \\ 
        233498.30       & C$_2$H$_5$CN  & $26_{5,21}-25_{5,20}$         & 178.9   & 371.1 & -2.98         & -                     & JPL           & 5.2$\pm$0.2     & 97.2$\pm$0.1     & 41.5$\pm$2.7 \\ 
        233553.24       & CH$_3$OCHO    & $19_{7,12}-18_{7,11}$ A       & 332.0   & 43.7  & -3.78         & -                     & JPL           & 5.8$\pm$0.4     & 97.1$\pm$0.1     & 20.1$\pm$2.2 \\ 
        233571.02       & C$_2$H$_5$OH  & $13_{5,8}-13_{4,9}$           & 107.9   & 12.8  & -4.15         & -                     & JPL           & 5.2$\pm$0.5     & 97.5$\pm$0.2     & 15.5$\pm$2.6 \\ 
        233632.27       & CH$_3$OCH$_3$& $25_{5,20}-25_{4,21}$  & 331.9 & 402.9   & -4.13         & -                     & CDMS          & 5.5$\pm$0.2     & 97.3$\pm$0.1     & 35.1$\pm$1.9 \\ 
        233649.88       & CH$_3$OCHO    & $19_{12,7}-18_{12,6}$ E       & 207.6   & 30.5  & -3.94         & -                     & JPL           & 4.0$\pm$0.1     & 97.2$\pm$0.1     & 19.6$\pm$0.7 \\ 
        233655.34       & CH$_3$OCHO    & $19_{12,7}-18_{12,6}$ A       & 207.6   & 30.5  & -3.94         & -                     & JPL           & 6.6$\pm$0.8     & 97.8$\pm$0.4     & 64.6$\pm$11.3 \\ 
        233670.98       & CH$_3$OCHO    & $19_{12,8}-18_{12,7}$ E       & 207.6   & 30.5  & -3.94         & -                     & JPL           & 4.6$\pm$0.3     & 97.1$\pm$0.1     & 19.5$\pm$2.2 \\ 
        233727.94       & CH$_3^{18}$OH & $5_{1,4}-4_{1,3}$ A           & 48.0    & 15.6  & -4.28         & 5.5(5) $^a$   & CDMS          & 4.5$\pm$0.4     & 97.3$\pm$0.2     & 14.8$\pm$2.3 \\ 
        233734.72       & HC(O)NH$_2$   & $11_{4,8}-10_{4,7}$           & 114.9   & 124.8 & -3.03         & -                     & CDMS          & 6.8$\pm$0.3     & 96.6$\pm$0.1     & 17.4$\pm$1.5 \\ 
        233745.61       & HC(O)NH$_2$   & $11_{4,7}-10_{4,6}$           & 114.9   & 124.8 & -3.03         & -                     & CDMS          & 5.3$\pm$0.4     & 97.5$\pm$0.1     & 13.5$\pm$1.6 \\ 
        233753.96       & CH$_3$OCHO    & $18_{4,14}-17_{4,13}$ E       & 114.4   & 45.8  & -3.74         & -                     & JPL           & 5.4$\pm$0.3     & 97.3$\pm$0.1     & 45.4$\pm$4.2 \\ 
        233777.52       & CH$_3$OCHO    & $18_{4,14}-17_{4,13}$ A       & 114.4   & 45.8  & -3.74         & -                     & JPL           & 4.2$\pm$0.2     & 97.1$\pm$0.1     & 35.3$\pm$3.0 \\ 
        233795.67       & CH$_3$OH              & $18_3- - 17_4-$                       & 446.59  & 5.5   & -4.66         & -                     & JPL           & 5.0$\pm$0.3             & 96.9$\pm$0.1          & 49.2$\pm$5.9 \\ 
        233845.23       & CH$_3$OCHO    & $19_{11,8}-18_{11,7}$ E       & 192.4   & 33.7  & -3.89         & -                     & JPL           & 5.4$\pm$0.2     & 97.3$\pm$0.1     & 26.4$\pm$1.6 \\ 
        233854.29       & CH$_3$OCHO    & $19_{11,9}-18_{11,8}$ A       & 192.4   & 33.7  & -3.89         & -                     & JPL           & 4.2$\pm$0.2     & 97.1$\pm$0.1     & 28.6$\pm$2.7 \\ 
        233867.19       & CH$_3$OCHO    & $19_{11,9}-18_{11,8}$ E       & 192.4   & 33.7  & -3.89         & -                     & JPL           & 4.3$\pm$0.4     & 96.9$\pm$0.2     & 20.6$\pm$3.3 \\ 
        233896.58       & HC(O)NH$_2$   & $11_{3,9}-10_{3,8}$           & 94.1    & 133.1 & -3.03         & -                     & CDMS          & 6.2$\pm$0.5     & 97.3$\pm$0.2     & 19.8$\pm$2.9 \\ 
        233916.95       & CH$_3$OH              & $13_3-        - 14_4-$                        & 868.5   & 2.5   & -5.26                 & -                     & JPL           & 4.9$\pm$0.4             & 97.1$\pm$0.2          & 13.3$\pm$2.0 \\ 
        233951.12       & C$_2$H$_5$OH  & $13_{5,9}-13_{4,10}$          & 107.9   & 12.8  & -4.15         & -                     & JPL           & 5.2$\pm$0.3     & 97.2$\pm$0.1     & 18.3$\pm$1.6 \\ 
        234011.58       & $^{13}$CH$_3$OH& $5_{1,5}-4_{1,4}$ ++         & 48.3    & 3.9   & -4.28         & 1.1(7) $^a$   & CDMS          & 4.7$\pm$0.2           & 96.9$\pm$0.1            & 35.3$\pm$2.3 \\ 
        234112.33       & CH$_3$OCHO    & $19_{10,9}-18_{10,8}$ E       & 178.5   & 36.6  & -3.85         & -                     & JPL           & 5.0$\pm$0.2     & 97.4$\pm$0.1     & 28.2$\pm$2.2 \\ 
        234124.88       & CH$_3$OCHO    & $19_{10,9}-18_{10,8}$ A       & 178.5   & 36.6  & -3.85         & -                     & JPL           & 4.7$\pm$0.3     & 96.9$\pm$0.1     & 35.5$\pm$4.3 \\ 
        234134.60       & CH$_3$OCHO    & $19_{10,10}-18_{10,9}$ E      & 178.5   & 36.6  & -3.85         & -                     & JPL           & 4.8$\pm$0.3     & 97.4$\pm$0.1     & 27.6$\pm$2.7 \\
        234255.16       & C$_2$H$_5$OH  & $12_{5,8}-12_{4,9}$           & 96.9    & 11.6  & -4.16         & -                     & JPL           & 4.6$\pm$0.4     & 97.1$\pm$0.2     & 12.2$\pm$2.0 \\ 
        234315.50       & HC(O)NH$_2$   & $11_{3,8}-10_{3,7}$           & 94.1    & 133.1 & -3.03         & -                     & CDMS          & 6.3$\pm$0.4     & 96.8$\pm$0.2     & 17.4$\pm$2.4 \\ 
        234336.11       & CH$_3$OCHO    & $19_{6,14}-18_{6,13}$ A       & 323.5   & 45.5  & -3.76         & -                     & JPL           & 4.6$\pm$0.4     & 97.2$\pm$0.2     & 14.5$\pm$2.2 \\ 
        234381.27       & CH$_3$OCHO    & $19_{5,15}-18_{5,14}$ A       & 316.4   & 46.9  & -3.74         & -                     & JPL           & 4.4$\pm$0.2     & 97.0$\pm$0.1     & 16.9$\pm$1.7 \\
        234406.45       & C$_2$H$_5$OH  & $11_{5,6}-11_{4,7}$           & 86.8    & 10.4  & -4.17         & -                     & JPL           & 4.4$\pm$0.1     & 97.2$\pm$0.1     & 11.2$\pm$0.7 \\ 
        234423.96       & C$_2$H$_5$CN  & $26_{4,22}-25_{4,21}$         & 169.1   & 376.3 & -2.98         & -                     & JPL           & 5.3$\pm$0.1     & 97.2$\pm$0.1     & 42.9$\pm$1.4 \\ 
        \hline
\end{tabular}

\medskip
\textit{Notes}: All observational values are integrated over the continuum core.

$^a$ Collision rate value from the main isotopologue. 

$^b$ The line is self-absorbed or filtered. The values correspond to the zero-base line width, the velocity of the line maximum, and the integration of the positive $T_{\rm mb}$ values.

\end{table}

\newpage
\begin{table}[p]                        
\setlength\tabcolsep{1mm}
\renewcommand\thetable{A\arabic{table}} 

\caption{Same as Table \ref{table:LinesC3} but towards core \#6.}
\setlength{\tabcolsep}{5pt}
\begin{tabular}{cccccccc|ccc}
        \hline\hline
        Frequency       & Molecule              & Transition                            & $E_{\rm up}$    &S$\mu^2$& log($A_{\rm ij}$)    & $n_{\rm crit}$        &Database       & Line width       & Velocity              & $\int{T_{\rm mb} d\varv}$      \\ 
        $$[MHz]         &                               & J=                                            & [K]             & [D$^2$]       & [s$^{-1}$]    & [cm$^{-3}$]   &                       & [km.s$^{-1}$]           & [km.s$^{-1}$]                 & [K.km.s$^{-1}$] \\
        \hline
        216109.78       & CH$_3$OCHO    & $19_{2,18}-18_{2,17}$ E       & 109.3   & 49.4  & -3.83         & -                     & JPL           & 3.2$\pm$0.9             & 95.5$\pm$0.4          & 2.1$\pm$0.5 \\
        216115.57       & CH$_3$OCHO    & $19_{2,18}-18_{2,17}$ A       & 109.3   & 49.4  & -3.83         & -                     & JPL           & 3.8$\pm$1.5             & 95.9$\pm$0.4          & 2.8$\pm$0.8 \\
        216147.36       & OC$^{33}$S            & $18-17$                               & 98.6    & 9.2   & -4.04                 & 1.3(6) $^a$   & CDMS          & 2.1$\pm$0.6             & 95.4$\pm$0.3          & 1.1$\pm$0.3   \\
        216210.91       & CH$_3$OCHO    & $19_{1,18}-18_{1,17}$ E       & 109.3   & 49.4  & -3.83         & -                     & JPL           & 1.8$\pm$0.4             & 95.1$\pm$0.2          & 1.8$\pm$0.4   \\
        216216.54       & CH$_3$OCHO    & $19_{1,18}-18_{1,17}$ A       & 109.3   & 49.4  & -3.83         & -                     & JPL           & 2.4$\pm$0.9             & 95.9$\pm$0.4          & 1.6$\pm$0.5           \\
        \hdashline
        217104.98       & SiO v=0               & $5-4$                                 & 31.3    & 48.0  & -3.28         & 4.8(6)                & CDMS          & 5.1$\pm$0.7             & 95.8$\pm$0.3          & 20.0$\pm$0.6          \\
        217238.54       & DCN                   & $3-2$                                 & 20.9    & 80.5  & -3.34                 & -                     &CDMS           & 1.7$\pm$0.3             & 95.6$\pm$0.1          & 3.6$\pm$0.4           \\
        \hdashline
        218199.00       & O$^{13}$CS            & $18-17$                               & 99.5    & 9.2   & -4.52         & 4.1(5) $^a$   & CDMS          & 2.1$\pm$0.4           & 95.1$\pm$0.2            & 2.0$\pm$0.3           \\
        218222.19       & H$_2$CO               & $3_{0,3}-2_{0,2}$                     & 21.0    & 16.3  & -3.55         & 2.6(6)                & CDMS          & 22 $^b$         & 94.8 $^b$                     & >45 $^b$              \\
        218280.90       & CH$_3$OCHO    & $17_{3,14}-16_{3,13}$ E       & 99.7    & 43.6  & -3.82         & -                     & JPL           & 2.8$\pm$0.5             & 96.1$\pm$0.2          & 2.0$\pm$0.4           \\
        218297.89       & CH$_3$OCHO    & $17_{3,14}-16_{3,13}$ A       & 99.7    & 43.6  & -3.82         & -                     & JPL           & 4.0$\pm$0.6             & 95.2$\pm$0.4          & 2.4$\pm$0.4   \\
        218324.72       & HC$_3$N               & $24-23$                               & 131.0   & 334.2 & -3.08         & 1.7(7)                & CDMS          & 4.4$\pm$0.5             & 96.3$\pm$0.2          & 10.7$\pm$0.9          \\
        \hdashline
        219560.35       & C$^{18}$O             & $2-1$                                 & 15.8    & 0.02  & -6.22         & 9.9(3) $^a$   & CDMS          & 10.5 $^b$             & 95.5 $^b$                       & >14 $^b$              \\
        \hdashline
        219908.53       & H$_2$$^{13}$CO& $3_{1,2}-2_{1,1}$                     & 32.9    & 43.5  & -3.59         & 2.3(6) $^a$   & CDMS          & 2.4$\pm$0.6           & 96.0$\pm$0.2            & 6.3$\pm$1.2           \\
        219949.44       & SO                    & $6_{5}-5_{4}$                 & 35.0    & 14.0  & -3.87         & 2.3(6)                & CDMS          & 3.5$\pm$0.3             & 96.1$\pm$0.1          & 18.3$\pm$1.0          \\
        \hdashline
        230315.79       & CH$_3$CHO     & $12_{2,11}-11_{2,10}$ E       & 81.1    & 147.4 & -3.38         & -                     & JPL           & 2.9$\pm$1.1             & 95.3$\pm$0.2          & 3.5$\pm$1.3           \\
        230538.00       & CO                    & $2-1$                                 & 16.6    & 0.02  & -6.16         & 1.1(4)                & CDMS          & -                               & -                                     & >25 $^b$ \\
        \hdashline
        231060.99       & OCS                   & $19-18$                               & 110.9   & 9.7   & -4.45         & 4.9(5)                & CDMS          & 3.7$\pm$0.2             & 95.7$\pm$0.1          & 11.6$\pm$0.4          \\
        231220.68       & $^{13}$CS             & $5-4$                                 & 33.3    & 38.3  & -3.60         & 4.3(6) $^a$   & JPL           & 3.4$\pm$0.3           & 95.4$\pm$0.1            & 6.6$\pm$0.5           \\
        231269.90       & CH$_3$CHO     & $12_{6,7}-11_{6,6}$ E         & 153.4   & 113.9 & -3.48         & -                     & JPL           & 3.3$\pm$1.0             & 96.4$\pm$0.3          & 2.7$\pm$0.6           \\
        231281.11       & CH$_3$OH              & $10_2$-$ - 9_3$-                      & 165.3   & 2.68  & -4.73         & 2.5(7)                & JPL           & 3.3$\pm$0.6             & 95.9$\pm$0.3          & 3.2$\pm$0.5           \\
        231467.50       & CH$_3$CHO     & $12_{4,8}-11_{4,7}$ E         & 108.4   & 135.0 & -3.41         & -                     & JPL           & 1.5$\pm$0.3             & 95.1$\pm$0.1          & 1.4$\pm$0.2           \\
        231506.29       & CH$_3$CHO     & $12_{4,9}-11_{4,8}$ E         & 108.2   & 135.0 & -3.41         & -                     & JPL           & 1.4$\pm$0.5             & 95.9$\pm$0.2          & 1.1$\pm$0.3           \\
        \hdashline
        231595.27       & CH$_3$CHO     & $12_{3,10}-11_{3,9}$ E        & 92.6    & 142.3 & -3.39         & -                     & JPL           & 3.7$\pm$0.9                     & 95.3$\pm$0.5                  & 1.4$\pm$0.4           \\
        231748.72       & CH$_3$CHO     & $12_{3,10}-11_{3,9}$ E        & 92.5    & 141.1 & -3.39         & -                     & JPL           & 2.9$\pm$1.1                     & 95.6$\pm$0.5                  & 1.4$\pm$0.4           \\
        231987.93       & CH$_3$OCH$_3$ & $13_{0,13}-12_{1,12}$ AE      & 80.9    & 101.9 & -4.04         & -                     & CDMS          & 2.3$\pm$0.7                     & 95.9$\pm$0.2                  & 1.5$\pm$0.4           \\
        232418.52       & CH$_3$OH              & $10_2$+$-9_3$+                        & 165.4   & 2.68  & -4.73         & 2.2(7)                & JPL           & 2.8$\pm$0.6                     & 96.2$\pm$0.3                  & 1.3$\pm$0.3           \\
        \hdashline
        232945.80       & CH$_3$OH              & $10_{-3}-11_{-2}$ E2          & 190.4   & 3.0   & -4.67         & 1.6(7)                & JPL           & 3.7$\pm$0.8             & 95.6$\pm$0.5          & 1.9$\pm$0.4           \\
        233226.79       & CH$_3$OCHO    & $19_{4,16}-18_{4,15}$ A       & 123.2   & 48.0  & -3.74         & -                     & JPL           & 2.9$\pm$1.0             & 95.4$\pm$0.6          & 1.8$\pm$0.6           \\
        233753.96       & CH$_3$OCHO    & $18_{4,14}-17_{4,13}$ E       & 114.4   & 45.8  & -3.74         & -                     & JPL           & 2.9$\pm$1.2             & 96.2$\pm$0.6          & 1.7$\pm$0.7           \\
        233777.52       & CH$_3$OCHO    & $18_{4,14}-17_{4,13}$ A       & 114.4   & 45.8  & -3.74         & -                     & JPL           & 2.9$\pm$0.8             & 96.1$\pm$0.4          & 2.3$\pm$0.6           \\
        234011.58       & $^{13}$CH$_3$OH& $5_{1,5}-4_{1,4}$ ++         & 48.3    & 3.9   & -4.28         & 1.1(7) $^a$   & CDMS          & 5.2$\pm$0.5           & 96.5$\pm$0.2            & 5.2$\pm$1.8           \\
        \hline
\end{tabular}

\medskip
\textit{Notes}: All observational values are integrated over the continuum core.

$^a$ Collision rate value from the main isotopologue. 

$^b$ The line is self-absorbed or filtered. The values correspond to the zero-base line width, the velocity of the line maximum, and the integration of the positive $T_{\rm mb}$ values.

\label{table:LinesC6}
\end{table}

\newpage

\begin{table}[htb]     
\renewcommand\thetable{A\arabic{table}}           
        \begin{center}
                \setlength\tabcolsep{3mm}
                \caption[]{Column densities in core \#3 estimated for a source size of 1000 au and a temperature of 190 K.}

                \begin{tabular}{lc} 
                \hline\hline
                Molecule        &$N$ [cm$^{-2}$]\\
                \hline  
                C$^{18}$O               & >8.4(17)      \\      
                CH$_3$OH                & 1.4(18)               \\
                $^{13}$CH$_3$OH & 2.8(17)               \\      
                CH$_3$CHO               & 3.8(16)               \\      
                $^{13}$CS               & 3.2(15)               \\      
                CH$_3$OCH$_3$   & 4.7(17)       \\      
                CH$_3$OCHO              & 3.8(17)       \\      
                OCS                             & 1.0(17)               \\      
                O$^{13}$CS              & 1.2(16)               \\      
                OC$^{33}$S              & 5.5(15)               \\      
                SiO v=0                 & >1.9(15)      \\
                SO                              & 1.5(16)               \\
                H$_2$CO                 & >2.0(16)      \\
                H$_2^{13}$CO    & 7.8(15)               \\
                HC$_3$N                 & 2.4(15)               \\
                DCN                             & 1.5(15)               \\
                CO                              & >2.1(19)      \\
                CH$_3^{18}$OH   & 8.2(16)       \\
                $^{13}$CH$_3$CN & 1.9(15)       \\
                HC(O)NH$_2$             & 3.9(15)       \\
                C$_2$H$_5$OH    & 7.2(16)       \\
                H$_2$C$^{34}$S  & 4.0(15)               \\
                CH$_3$COCH$_3$  & 9.0(16)               \\
                C$_2$H$_3$CN    & 8.3(15)       \\
                C$_2$H$_5$CN    & 1.7(16)       \\
                \hline
                \multicolumn{2}{l}{Column densities estimated from the isotopologues}\\
                CH$_3$OH                & 2.0(19)       \\
                CS                              & 1.4(17)       \\
                OCS                             & 5.2(17)       \\
                H$_2$CO                 & 3.5(17)       \\
                CO                              & >3.8(19)      \\
                \hline
                \label{table:Ndiffscenario3}
                \end{tabular}
        \end{center}
\textit{Note}: $a(b)=a\times10^b$.
\end{table}

\newpage
\begin{figure*}[t!]
\renewcommand\thefigure{A\arabic{figure}}
\setcounter{figure}{0} 
        \centering
        \includegraphics[width=0.98\linewidth]{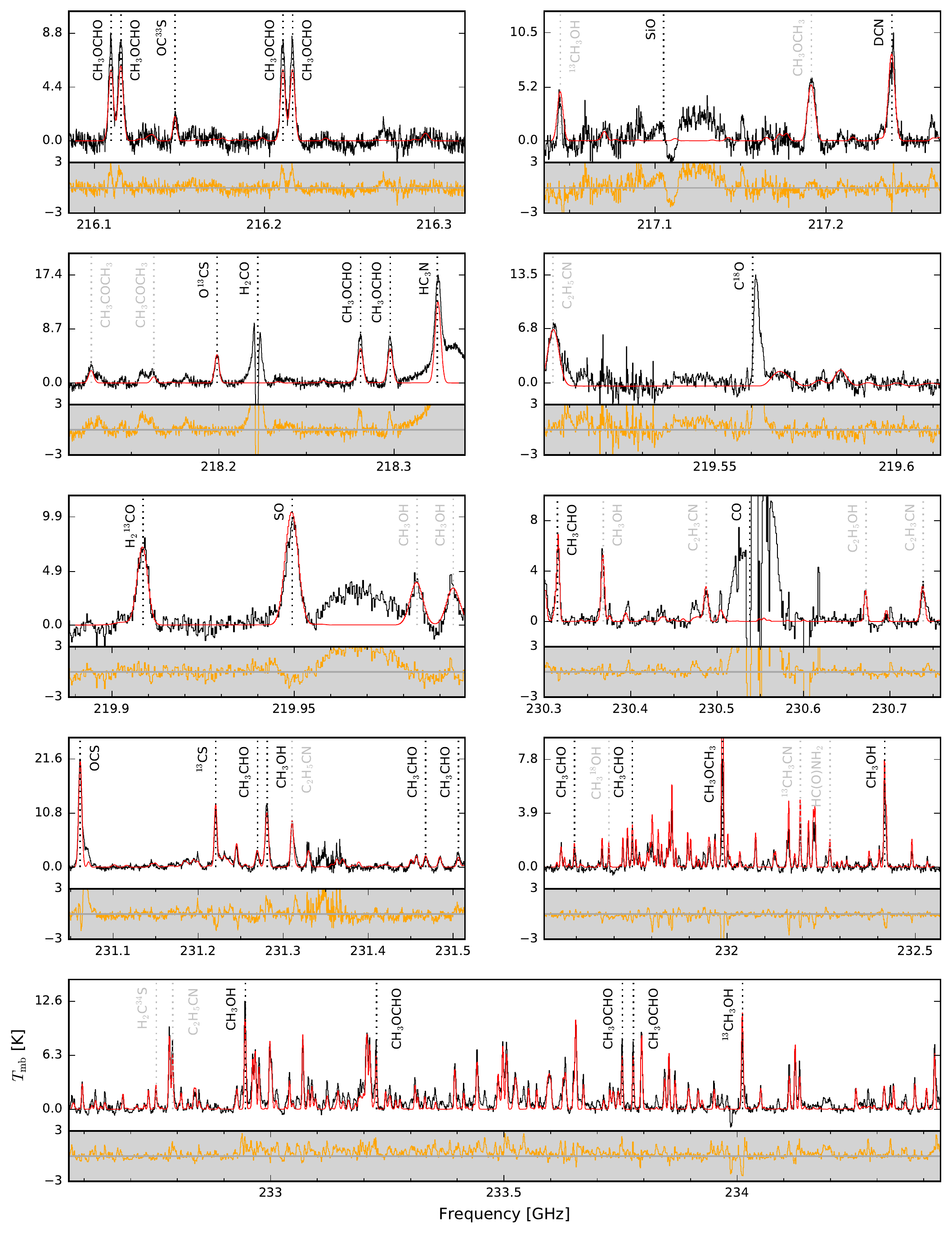}
        \caption{Observed spectra (in black), synthetic spectra (in red) and residuals (in orange) towards core \#3 for all the ALMA bands. The observed spectra are integrated over the continuum cores (see Table \ref{table:sources}). All the molecular lines detected in core \#6 (see Table \ref{table:LinesC6}) are indicated by black dotted lines and some of the additional molecular lines detected in core \#3 are indicated in grey. The parameters used for the synthetic spectra are listed in Table \ref{table:MoleculeProp}. The Cycle 3 spectral band Cont. 1, which overlaps with Cycle 2 data Cont. 2, is ignored here.}
        \label{fig:Spectra-core3}
\end{figure*}

\newpage

\begin{figure*}[t!]
\renewcommand\thefigure{A\arabic{figure}}

        \centering
        \includegraphics[width=0.98\linewidth]{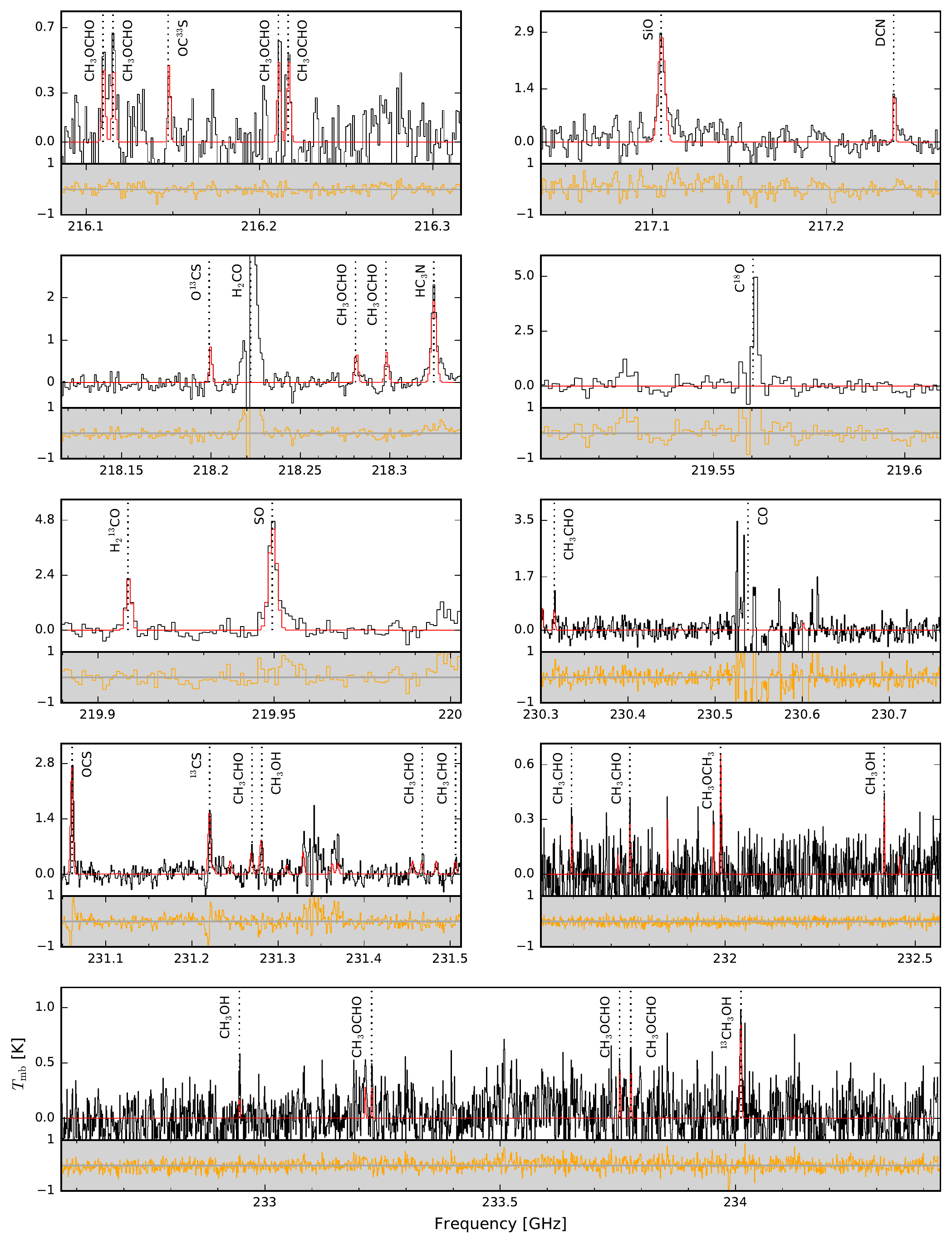}
        \caption{Same as Fig. \ref{fig:Spectra-core3} but for source \#6. The spectra are Hanning smoothed to the same frequency resolution, $\Delta f = 0.976$ MHz. The synthetic spectra are calculated for a temperature of 30 K, which corresponds to the very beginning of the protostellar phase.}
        \label{fig:Spectra-core6}
\end{figure*}

\end{document}